\RequirePackage{fix-cm}

\documentclass[smallextended]{svjour3modified}       

\smartqed  
\usepackage{graphicx}
\usepackage{lmodern}
\usepackage{mathtools}
\usepackage{bbm}
\usepackage{physics}
\usepackage{color}
\usepackage{ifthen}

\newcommand{\K}[1]{\ensuremath{\left(#1\right)}}
\newcommand{\Ke}[1]{\ensuremath{\left[#1\right]}}

\renewcommand{\v}[1]{\ensuremath{\boldsymbol{#1}}}
\newcommand{\vh}[1]{\v{\hat{#1}}}

\newcommand{\mprescript}[3]{{{\vphantom{#3}}}^{#1}_{#2}\! #3 }

\newcommand{\ibraket}[4]{ \mprescript{}{#1}{ \braket{#2}{#3}_{#4}^{} } } 
\newcommand{\ibra}[2]{ \mprescript{}{#1}{ \bra{#2} } }
\newcommand{\iket}[2]{ \ket{#1}_{#2}^{} } 
\newcommand{\iketbra}[3]{
  \ifthenelse{\equal{#3}{}}{
    \iket{#1}{#2} \ibra{#2}{#1}
  }
  {
    \iket{#1}{#2} \ibra{#2}{#3}
  }
}

\newcommand{\imel}[5]{ \mprescript{}{#1}{ \mel{#2}{#3}{#4}_{#5}^{} } }

\newcommand{\p}[1][]{p^{#1}}
\newcommand{\q}[1][]{q^{#1}}
\newcommand{\pp}[1][]{p^{\prime #1}}
\newcommand{\qp}[1][]{q^{\prime #1}}

\newcommand{\qt}{\tilde{q}}
\newcommand{\ptp}[1][]{\tilde{p}^{\prime #1}}
\newcommand{\qtp}[1][]{\tilde{q}^{\prime #1}}

\newcommand{\rint}[1]{\int \dd{#1[]} #1[2]}

\newcommand{\angint}[1]{\int \dd{\Omega_{\v{#1}}}}

\newcommand{\reg}[2]{g_{l_#1}{\K{#2}}}

\newcommand{\gz}[2]{G_0^{\K{#1}}{\K{#2}}}

\newcommand{\rtm}[3]{\tau_{#1}{\K{#2;#3}}}

\newcommand{\y}[3]{Y_{#1,#2}{\K{#3}}}

\newcommand{\de}[1]{\delta{\K{#1}}}
\newcommand{\dt}[1]{\delta^{(3)}{\K{#1}}}
\newcommand{\da}[1]{\delta^{(\Omega)}{\K{#1}}}

\newcommand{\cgc}[2]{C{\K{#1 \,\vline \, #2}}}

\newcommand{\pmo}{\mathcal{P}_{nn}}
\newcommand{\pmospatial}{\mathcal{P}_{nn}^\mathrm{(spatial)}}
\newcommand{\pmospin}{\mathcal{P}_{nn}^\mathrm{(spin)}}

\newcommand{\pfv}[2]{\v{\pi}_{#1}{\K{#2}}}

\newcommand{\cy}[3]{\mathcal{Y}_{#1}^{#2}{\K{#3}}}
\newcommand{\cywa}[2]{\mathcal{Y}_{#1}^{#2}}

\newcommand{\ci}{\mathrm{i}}

\newcommand{\kcnp}{\kappa_{cnp}}
\newcommand{\kcnpv}{\v{\kappa}_{cnp}}
\newcommand{\kcnph}{\v{\hat{\kappa}}_{cnp}}
\newcommand{\kcnq}{\kappa_{cnq}}
\newcommand{\kcnqv}{\v{\kappa}_{cnq}}
\newcommand{\kcnqh}{\v{\hat{\kappa}}_{cnq}}

\newcommand{\kcnpp}{\kappa_{cnp}^\prime}
\newcommand{\kcnppv}{\v{\kappa}_{cnp}^{\prime}}
\newcommand{\kcnpph}{\v{\hat{\kappa}}_{cnp}^{\prime}}
\newcommand{\kcnqp}{\kappa^\prime_{cnq}}
\newcommand{\kcnqpv}{\v{\kappa}_{cnq}^{\prime}}
\newcommand{\kcnqph}{\v{\hat{\kappa}}_{cnq}^{\prime}}

\newcommand{\kncp}{\kappa_{ncp}}
\newcommand{\kncpv}{\v{\kappa}_{ncp}}

\newcommand{\kncq}{\kappa_{ncq}}
\newcommand{\kncqv}{\v{\kappa}_{ncq}}

\newcommand{\knnpv}{\v{\kappa}_{nnp}}

\newcommand{\knnqv}{\v{\kappa}_{nnq}}

\newcommand{\knnpp}{\kappa_{nnp}^\prime}
\newcommand{\knnppv}{\v{\kappa}_{nnp}^{\prime}}
\newcommand{\knnpph}{\v{\hat{\kappa}}_{nnp}^{\prime}}
\newcommand{\knnqp}{\kappa^\prime_{nnq}}
\newcommand{\knnqpv}{\v{\kappa}_{nnq}^{\prime}}
\newcommand{\knnqph}{\v{\hat{\kappa}}_{nnq}^{\prime}}

\newcommand{\oh}{\frac{1}{2}}

\newcommand{\id}{\mathbbm{1}}

\newcommand{\spin}{{\K{\mathrm{spin}}}} 

\newcommand{\spid}{\mathbbm{1}^\spin}

\newcommand{\kd}[2]{\delta_{#1, #2}}

\newcommand{\spiso}[2]{\sum_{#1=0}^1 \sum_{#2=-#1}^{#1} }

\newcommand{\hesix}{\textsuperscript{6}He}

\newcommand{\vpq}{\v{p},\v{q}}
\newcommand{\vpqh}{\vh{p},\vh{q}}
\newcommand{\vpqp}{\v{\pp},\v{\qp}}
\newcommand{\vpqph}{\vh{\pp},\vh{\qp}}

\newcommand{\pif}[2]{\v{\pi}_{#1}{\K{#2}}}
\newcommand{\pifnv}[2]{\pi_{#1}{\K{#2}}}
\newcommand{\pifvh}[2]{\vh{\pi}_{#1}{\K{#2}}}

\newcommand{\np}{n^\prime}

\newcommand{\flb}{\quad}
\newcommand{\flbc}{\quad \cross}

\newcommand{\bn}{\beta_0}

\newcommand{\bo}{\beta_1}

\newcommand{\LScoupling}{\(\v{L}\v{S}\)-coupling}
\newcommand{\jJcoupling}{\(\v{j}\v{J}\)-coupling}

\newcommand{\rhos}[2]{\ensuremath{
  \rho_{#1}{\K{#2}}
}}

\newcommand{\rhot}[2]{\ensuremath{
  \tilde{\rho}_{#1}{\K{#2}}
}}

\newcommand{\rhopw}[3]{\ensuremath{
  \rho_{#1}^{\K{#2}}{\K{#3}}
}}

\newcommand{\rhotpw}[3]{\ensuremath{
  \tilde{\rho}_{#1}^{\K{#2}}{\K{#3}}
}}

\newcommand{\stdff}{Heaviside form factor}

\newcommand{\ca}{c}

\newcommand{\ya}{\Upsilon}

\let\myhat\hat
\renewcommand{\hat}[1]{#1}

\begin{document}

\title{Momentum-space probability density of ${}^6$He in Halo Effective Field Theory\thanks{In memory of Ludwig Faddeev}}

\author{Matthias G\"obel \and
        Hans-Werner Hammer \and
        Chen Ji \and
        Daniel R.~Phillips
}

\institute{Matthias G\"obel \at
  Institut f\"ur Kernphysik, Technische Universit\"at Darmstadt, 64289 Darmstadt, Germany\\
  \email{goebel@theorie.ikp.physik.tu-darmstadt.de}
  \and
  Hans-Werner Hammer \at
  Institut f\"ur Kernphysik, Technische Universit\"at Darmstadt, 64289 Darmstadt, Germany\\
  ExtreMe Matter Institute EMMI, GSI Helmholtzzentrum f{\"u}r Schwerionenforschung GmbH, 64291 Darmstadt, Germany\\
  \email{Hans-Werner.Hammer@physik.tu-darmstadt.de}
  \and
 	Chen Ji \at
  Key Laboratory of Quark and Lepton Physics (MOE) and Institute of Particle Physics,
  Central China Normal University, Wuhan 430079, China\\
	\email{jichen@mail.ccnu.edu.cn}
  \and
  Corresponding author: Daniel R. Phillips \at
  Institute of Nuclear and Particle Physics and Department of Physics and
  Astronomy, Ohio University, Athens, OH 45701, USA\\
  Institut f\"ur Kernphysik, Technische Universit\"at Darmstadt, 64289 Darmstadt, Germany\\
  ExtreMe Matter Institute EMMI, GSI Helmholtzzentrum f{\"u}r Schwerionenforschung GmbH, 64291 Darmstadt, Germany\\
  \email{phillid1@ohio.edu}
}

\date{September 12, 2019}

\maketitle

\begin{abstract}
We compute the momentum-space probability density of ${}^6$He
at leading order in Halo EFT. In this framework,
the ${}^6$He nucleus is treated as a three-body problem with
a ${}^4$He core (\(\ca\)) and two valence neutrons (\(n\)). This
requires the \(nn\) and \(n \ca\) t-matrices as well as a
\(\ca nn\) force  as input in the Faddeev equations.
Since the \(n \ca\) t-matrix corresponds to an energy-dependent potential,
we consider the consequent modifications to the standard normalization
and orthogonality conditions. We find that these are small for momenta
within the domain of validity of Halo EFT. In this regime, the ${}^6$He
probability density is regulator independent, provided the cutoff is significantly above the
EFT breakdown scale.

\keywords{Effective field theory \and Energy-dependent potentials \and Two-neutron halos}
\end{abstract}

\section{Introduction}
\label{sec:intro}
Probability densities contain important information on the structure
of quantum mechanical objects. In chemistry, for example, the
atomic orbital model provides a means to visualize 
the electronic cloud.  In nuclear physics, it has long been 
held that wave functions can be measured in knock-out experiments as
long as the quasielastic approximation holds (see, e.g.,
Refs.~\cite{Ulrych:1998nk,Yaron:2002nv,Benhar:2006wy} and references
therein). However, in general
kinematic situations this is not
true: at shorter distance scales
strength can be shifted
between the initial-state wave function, the current operator,
and final-state interactions by means of unitary transformations
or a redefinition of field variables~\cite{Haag:1958vt,Kamefuchi:1961sb,Chisholm:1961tha,Furnstahl:2001xq,More:2017syr}.

Here we consider probability densities for halo nuclei.
Halo nuclei are characterized by a tightly bound core and a few
loosely bound valence nucleons. They show universal properties independent of
the details of their structure at short
distances~\cite{Jensen:2004zz,Braaten:2004rn}.
Due to this separation of scales in terms of a core momentum scale, $M_\mathrm{core}$,
and a halo momentum scale, $M_\mathrm{halo}\ll M_\mathrm{core}$,
halo nuclei can be described as effective
few-body systems in an expansion in
$M_\mathrm{halo}/M_\mathrm{core}$. This expansion is conveniently
implemented using the framework of Halo Effective Field Theory
(Halo EFT)~\cite{bertulani02,bedaque03,hammer17}.

Halo nuclei consisting of two valence neutrons and a core are an effective three-body problem
in Halo EFT.
In this case, the full wave function depends on two
Jacobi momenta. As we will discuss below this leads to some
complications in the definition of the probability density
due to the angular-momentum recoupling that arises in transformations
between
different sets of Jacobi coordinates.
These complications occur for any three-body
bound state in which interactions are defined on a partial-wave basis.

Here we focus on the momentum-space probability density for
${}^6$He, where two valence
neutrons interact with each other and with an $\alpha$-particle core.
$^6$He has
already been discussed at leading order in Halo EFT~\cite{Rotureau:2012yu,ji14}.
We follow the Halo EFT treatment of $^6$He
by Ji {\it et al.}~\cite{ji14},
which includes the ${}^1S_0$ \(nn\)
interaction and the ${}^2P_{3/2}$ \(n\ca\) interaction at leading
order using a dibaryon formalism. The
$\ca n$ t-matrix then corresponds to an energy-dependent potential.
This necessitates modification of the
standard normalization and orthogonality
conditions, which in turn affects the expression for the momentum-space probability densities.

Our approach is to construct two-body potentials, and corresponding amplitudes, 
that reproduce Halo EFT---or, equivalently, the effective-range expansion (ERE)---up to a given order and then solve the momentum-space Faddev equations.
By doing this we obtain a solution to the three-body problem in which the asymptotic
wave function is factorized and incorporates two-body subsystems with phase shifts described by the ERE.
But this is not the only way to achieve that end. 
In Refs. \cite{Fedorov_2001,Fedorov:2001wj} Fedorov and Jensen used the adiabatic hyperspherical approach and imposed 
short-distance boundary conditions on each Faddev component of the wave function that ensure
that two-body amplitudes agree with the ERE up to terms of \(\mathcal{O}{\K{k^4}}\).

Note that we refer to our approach to $^6$He as Halo Effective Field Theory (Halo
EFT), even though the wave function of $^6$He is obtained by solving the
Schr\"odinger/Faddeev equation for suitably chosen effective potentials.
Our approach has all the features of  an EFT: it provides a power-counting scheme to calculate higher-order corrections and
so offers a way to systematically improve calculations and obtain error estimates.
Many-body forces and other higher-order terms enter naturally as
specified by the counting scheme.
Consequently, it is common practice to refer to this framework
as ``Halo EFT": this nomenclature was established already in the paper of Ji {\it et al.} that computed ${}^6$He in this manner~\cite{ji14}.

The paper is organized as follows:
In Sec. \ref{sec:6HeHaloEFTreview} the Halo EFT for \hesix~is reviewed.
We write down the Faddeev equations for this system and compute the potentials necessary
to reproduce the leading-order $\ca n$ and $nn$ amplitudes in Halo EFT. This is done using
separable potentials with two different form-factor choices.
The coupling strengths thereby obtained are, in general, energy dependent.
The quantum mechanics of energy-dependent potentials is then discussed in Sec. \ref{sec:energydeppotentials}
at a generic level.
As an example the momentum-space probability density of a two-body bound state with an energy-dependent
contact interaction is worked out.
In Sec.~\ref{sec:probdensity}
we derive the expressions for the probability density of \hesix~with and without
the modifications due to energy dependence of the potential.
We show results for different regulator parameters and different form factors and observe that the low-momentum
part of the density is independent of these choices.
Furthermore, we show that the modifications of the probability density due to the energy dependence of the potentials
are small for momenta in the domain of validity of the EFT.
Finally, in Sec. \ref{sec:future} we summarize our results and give an outlook.
%
\section{Halo EFT for ${}^6$He}
\label{sec:6HeHaloEFTreview}
\subsection{From Halo EFT to the ${}^6$He wave function via the Faddeev equations}\label{subsec:heft}
Our aim is to solve the stationary three-body Schr\"odinger equation \(H\ket{\Psi} = E_3 \ket{\Psi}\) using the Faddeev equations, with the Hamiltonian \(H\) derived from the leading-order Halo EFT for \hesix.
In Halo EFT---as in the cluster models that came before it, e.g., Refs. \cite{Hebach:1967bpg,Shah:1970wu,Ghovanlou:1974zza,Chulkov:1990ac,Zhukov:1993aw}---\hesix~is a three-body problem with the neutrons \(n\) and the core \(c\) as degrees of freedom.
Within this review we follow \cite{ji14}, where Halo EFT for \hesix~is set up, the Faddeev amplitudes are calculated and the three-body system is renormalized.
A general discussion of the Faddeev equations can be found in \cite{gloeckle83} and a general review in the context of Halo EFT in \cite{hammer17}.
The path from the EFT to our Hamiltonian is that
the EFT's power counting determines
the t-matrices that describe two-body scattering to a given accuracy.
From there we infer the potential terms \(V\) of the Hamiltonian via the Lippmann-Schwinger equation.

In fact, for solving the Faddeev equations explicit expressions for the t-matrices are sufficient.
Expressions for the potentials are not necessary.
Nevertheless, there are two reasons that make explicit formulas for the potentials desirable.
The first, somewhat usual, reason is that explicit expressions for the potentials allow us to cross-check
whether the found solution really solves the stationary Schr\"odinger equation
\begin{equation}
\left(H_0 + \sum_i V_i+ V_3\right) \ket{\Psi} = E_3 \ket{\Psi}.
\label{eq:3BSE}
\end{equation}
Note that here we follow the usual spectator notation for the three-body problem, as inspired by Faddeev's treatment~\cite{Faddeev:1960su}, and denote the
kinetic energy operator by \(H_0\), the two-body potential corresponding to spectator $i$ by $V_i$, and the three-body potential by $V_3$.
The second reason is that when either the two-body or three-body potential in the Hamiltonian is energy dependent, the standard formula for wave function normalization
needs to be modified by adding extra terms related to the potential.

We start down the path to solution of Eq.~(\ref{eq:3BSE}) by reviewing the connection between the two-body t-matrices \(t\) and potentials \(V\).
It is given by the Lippmann-Schwinger equation for the t-matrix
\begin{equation}\label{eq:ls_op}
  t = V + V G_0 t\,,
\end{equation}
where \(G_0 = \K{E - H_0}^{-1}\) is the free Green's function.
In order to distinguish the different two-body interactions we introduce spectator indices \(i \in \{n, \np, c\}\).
The t-matrix \(t_i\) is the one corresponding to \(V_i\) via Eq.~(\ref{eq:ls_op}).
In the case of two-body potentials \(V_i\) we are limiting ourselves to separable potentials.
They have the advantage of significantly reducing the numerical effort needed to solve the Faddeev equations.
In momentum space a separable potential is given by
\begin{equation}\label{eq:def_v_sep}
  \mel{\v{p}}{V_i(E)}{\v{\pp}} = \K{2l_i+1} P_{l_i}{\K{\v{p}\cdot \v{\pp}}} \reg{i}{p} \lambda_i{\K{E}} \reg{i}{\pp}\,.
\end{equation}
Since  \(P_l\) denotes the \(l\)-th Legendre polynomial Eq.~(\ref{eq:def_v_sep}) encodes an interaction that only happens in the partial wave \(l_i\).
The strength of the potential is given by \(\lambda_i\), which can be energy dependent.
The dependence on off-shell momenta completely resides in the so-called form-factors \(g{\K{p}}\).
These form factors describe the shape of the potentials, their specific forms at high momenta should not affect low-energy observables. In an EFT treatment $\reg{i}{p}$ can be used to regularize ultraviolet divergences in scattering equations.
The separability of the potential implies also the separability of the t-matrix:
\begin{equation}
  \mel{\v{p}}{t_i(E)}{\v{\pp}} = \K{2l_i+1} P_{l_i}{\K{\v{p}\cdot \v{\pp}}} \reg{i}{p} \tau_i{\K{E}} \reg{i}{\pp}\,.
  \label{eq:sept}
\end{equation}
In our EFT treatment the expressions for the t-matrices are given according to a power counting in scattering parameters and/or binding momenta. They are also defined in terms of these, effective-range expansion, parameters.  $\tau_i{\K{E}}$ is therefore determined once the power counting and order of the calculation have been specified.
The strength parameter \(\lambda_i\) follows by partial-wave projecting and reshaping the Lippmann-Schwinger equation (see, e.g., Ref.~\cite{Afnan:1977pi}):
\begin{equation}\label{eq:lb_ls}
  \lambda_i^{-1}{\K{E}} = \tau_i^{-1}{\K{E}} + 4\pi \int_0^\infty \frac{q^2 \K{ \reg{i}{q} }^2 }{ E - \frac{q^2}{2\mu_{jk}} + \mathrm{i}\epsilon} \dd{q}\,.
\end{equation}

Now that we have summarized the two-body sector
we can proceed to the solution of the three-body problem using the Faddeev equations.
In a first step we neglect three-body forces, they will be included later.
Our aim is to solve the stationary Schr\"odinger equation \eqref{eq:3BSE}.
Since the system is bound, we specify \(E_3 = - B_3^{\K{0}}\), with \(B_3^{\K{0}}\) the binding energy of the three-body system relative to the $\ca nn$ threshold.
Following Faddeev the total state is decomposed into the Faddeev components
\begin{equation}\label{eq:def_fd_wfc}
  \ket{\psi_i} \coloneqq G_0 V_i \ket{\Psi}\,,
\end{equation}
which fulfill
\begin{equation}\label{eq:fd_dcmp}
  \sum_i \ket{\psi_i} = \ket{\Psi}\,.
\end{equation}
The Schr\"odinger equation then becomes equivalent to the set of coupled Faddeev equations:
\begin{equation}\label{eq:fd}
  \ket{\psi_i} = G_0 t_i \sum_{j \neq i} \ket{\psi_j}\,.
\end{equation}

At this point it is useful to discuss the basis states in terms of which the obtained result \(\ket{\Psi}\) shall be represented.
In the center-of-mass frame two independent momentum variables are needed to describe the state of the three-particle system.
We use standard Jacobi coordinates \(\v{p}\) and \(\v{q}\), where \(\v{p}\) is the relative momentum in the two-body subsystem.
The momentum of the third particle relative to the center of mass of the two-body subsystem is given by \(\v{q}\).
Depending on the choice of the third, spectator, particle
three sets of Jacobi coordinates can be defined.
The relations between the Jacobi momenta and the single particle momenta \(k_i\) are given by
\begin{align}
  \v{p}_i &\coloneqq \mu_{jk} \K{\frac{\v{k}_j}{m_j} - \frac{\v{k}_k}{m_k}}\,, \\
  \v{q}_i &\coloneqq \mu_{i\K{jk}} \K{\frac{\v{k}_i}{m_i} - \frac{\v{k}_j + \v{k}_k}{M_{jk}}}\,,
\end{align}
where \(M_{jk} \coloneqq m_j + m_k\), with reduced masses \(\mu_{jk} \coloneqq m_j m_k / \K{m_j + m_k}\) and
\(\mu_{i\K{jk}} \coloneqq m_i M_{jk} / \K{m_i + M_{jk}}\). Different sets of Jacobi momenta with other spectators are given by cyclic permutations of $ijk$.

We now discuss the description of angular momenta.
The relative orbital angular momentum of the two-body subsystem is given by \(\v{l}\), the one of the third particle with
respect to the two-body subsystem is denoted by \(\v{\lambda}\).
The total spin of the two-body subsystem is given by \(\v{s}\), the spin of the third particle by \(\v{\sigma}\).
The total angular momentum of the three-body system is \(\v{J}\).
In order to do the angular momentum coupling one has the choice between \jJcoupling~or \LScoupling.
In \jJcoupling~the angular momenta of the subsystem are coupled to \(\v{j} \coloneqq \v{l} + \v{s}\).
The angular momenta of the third particle make up \(\v{I} \coloneqq \v{\lambda} + \v{\sigma}\).
Often a description of the states in the \(\v{j}\v{J}\)-coupling scheme is useful (see \cite{gloeckle83}):
\begin{align}\label{eq:jJ_coupling}
  \ket{\K{l,s}j, \K{\lambda, \sigma} I;J,M} \coloneqq \sum_{\mathclap{m_j+m_I=M}} \cgc{j m_j I m_I}{J M} \ket{l,s;j, m_j} \ket{\lambda,\sigma;I, m_I \vphantom{l,j}}\,,
\end{align}
where \(\cgc{l_1 m_1 l_2 m_2}{L M} \coloneqq \braket{l_1 m_1 l_2 m_2}{(l_1 l_2) L M}\) is the Clebsch-Gordan coefficient.
In the case of \hesix~the quantum numbers of the ground state \(J^{\pi} = 0^+\). Our assumptions, that the \(nn\) interaction happens in the \({}^1S_0\) channel
and the \(nc\) one acts in the \({}^{2}P_{3/2}\), then fix all quantum numbers in the case of \jJcoupling.
This is summarized in the multiindices
\begin{align}
  \ket{\Omega_c} &\coloneqq \ket{\K{0,0}0, \K{0, 0} 0;0,0}\,, \\
  \ket{\Omega_n} &\coloneqq \ket{\K{1,\oh}\frac{3}{2}, \K{1, \oh} \frac{3}{2};0,0}\,,
\end{align}
where the quantum numbers are written in the same order as in \eqref{eq:jJ_coupling} and the multiindex $\ket{\Omega_i}$ is implicitly defined to mean that those quantum numbers refer to the co-ordinate system in which particle $i$ is the spectator.
The representations of these indices in \LScoupling~are given in \cite{ji14}.

Now we are equipped to set up the Faddeev equations.
We insert the identity \( \id = \sum_\Omega \rint{\p} \rint{\q} \ketbra{p,q;\Omega}\) into Eq. (\ref{eq:fd}) in order to
obtain a representation of the Faddeev equations.
For this purpose the matrix elements of the t-matrix and the Green's function are necessary:
\begin{align}
    \imel{i}{p,q;\Omega}{t_i{\K{E_3}}}{\pp,\qp;\Omega^\prime}{i} &= 4\pi \reg{i}{p} \tau_i{\left(E_3 - \frac{q^2}{2 \mu_{i\K{jk}}}\right)} \reg{i}{\pp} \label{eq:t_pw_me} \nonumber \\
      & \flbc \kd{\Omega}{\Omega^\prime} \kd{\Omega}{\Omega_i} \frac{\de{q-\qp}}{q^2}\,, \\
    \imel{i}{p,q;\Omega}{G_0\K{E_3}}{\pp,\qp;\Omega^\prime}{i} &= \K{E_3-\frac{p^2}{2\mu_{jk}}-\frac{q^2}{2\mu_{i\K{jk}}} }^{-1} \nonumber \\
      & \flbc \kd{\Omega}{\Omega^\prime} \frac{\de{p-\pp}}{p^2} \frac{\de{q-\qp}}{q^2}\,,
\end{align}
where \(E_3\) is the energy of the three-particle system.

In the case of a separable potential it is useful to define new components \(\ket{F_i}\),
which lead to a simpler set of equations~\cite{Afnan:1977pi}.
These are given by
\begin{equation}\label{eq:def_fdt}
  \ket{\psi_i} \eqqcolon G_0 t_i \ket{F_i}\,.
\end{equation}
Now, the Faddeev equations become
\begin{equation}\label{eq:fd_f_g}
  \ket{F_i} = \sum_{j \neq i} G_0 t_j \ket{F_j}\,.
\end{equation}
By inserting identities and some reshaping one obtains the final version of the Faddeev
equations:
\begin{equation}\label{eq:fd_f}
  F_i{\K{q}} = 4 \pi \sum_{j \neq i} \rint{\qp} X_{ij}{\K{q,\qp;E_3}} \tau_j{\K{\qp;E_3}} F_j{\K{\qp}}\,,
\end{equation}
with the definitions
\begin{align}
  F_i{\K{q}} &\coloneqq \rint{\p} \reg{i}{p} \ibraket{i}{p,q;\Omega_i}{F_i}{}\,, \\
  X_{ij}{\K{q,\qp;E_3}} &\coloneqq \rint{\p} \reg{i}{p} G_0^{\K{i}}{\K{p,q;E_3}} \nonumber \\
  & \flbc \rint{\pp} \ibraket{i}{p,q;\Omega_i}{\pp,\qp;\Omega_j}{j} \reg{j}{\pp} \,.
\end{align}
Eq.~(\ref{eq:fd_f}) is a set of homogeneous Fredholm integral equations of the second kind.
It can be solved numerically by discretization of the momentum-space integral.
Eqs.~(\ref{eq:def_fdt}) and (\ref{eq:fd_dcmp}) then determine how the wave function is obtained
from the solution for \(F_i{\K{q}}\).

The two integrals in \(X_{ij}{\K{q,\qp}}\) can be reduced to one integration over an angle using the delta functions in the
recoupling coefficient $\ibraket{i}{p,q;\Omega_i}{\pp,\qp;\Omega_j}{j}$.
In the case of form factors of the type \(g_l{\K{p}} = p^l \theta{\K{\beta_l -p }}\) this integral can then be evaluated analytically
to yield Legendre functions of the second kind if we neglect the Heaviside step function \(\theta\).
However, here we carry out this last integral numerically, since this makes it possible to extend the calculation to different form factors at a low effort.
By making some adjustments the additional numerical cost can be kept small.
This has the additional advantage that no discrepancies due to neglecting the step function are introduced.
Checks of the obtained solution are therefore better fulfilled.
The formulas actually used for \(X_{ij}\) are given in appendix~\ref{subsec:num_Xij}.

\hesix~contains two neutrons outside the $\alpha$ core, and the state \(\ket{\Psi}\) has to be antisymmetrized with respect to their exchange.
Demanding \(-\pmo \ket{\Psi} = \ket{\Psi}\), where \(\pmo\) is the \(nn\) permutation operator, and stating \(V_{\np} = \K{-\pmo} V_n \K{-\pmo}\), one
finds (using \([\pmo,V_c]=0\))
that \(\ket{\psi_{\np}} = - \pmo \ket{\psi_n}\).
Furthermore \(\ket{F_{\np}} = - \pmo \ket{F_n}\) holds. So, in
both versions of the Faddeev equations---\eqref{eq:fd} and \eqref{eq:fd_f_g}---
only two out of three equations are linearly independent.

Now that the solution procedure has been discussed we specify the t-matrices at leading order in Halo EFT:
\begin{align}
\langle \v{p}|t_{c}(E)|\v{p}' \rangle&= \frac{1}{4\pi^2 \mu_{nn}} \frac{1}{\gamma_0 + \mathrm{i} k}\,, \label{eq:t_nn}\\
\langle \v{p}|t_{n}(E)|\v{p}' \rangle&= \frac{3 \v{p} \v{p}'}{4\pi^2 \mu_{nc}} \frac{1}{\gamma_1\K{k^2-k_R^2}}\,,\label{eq:t_nc}
\end{align}
where \(k = \sqrt{2\mu E}\) is the on-shell momentum of the corresponding system and we have assumed that \(p\) and \(p^\prime\) are much smaller than the cutoff.
For a scattering volume \(a_1\)
and \(p\)-wave effective range \(r_1\) the momentum of the resonance in the \(nc\) interaction is given by \(k_R = \sqrt{2/\K{a_1 r_1}}\).
Furthermore, we designate \(\gamma_1 = -r_1/2\) and \(\gamma_0 = 1/a_0\) where \(a_0\) is the \(s\)-wave scattering length. There is then a pole in the \(s\)-wave at \(k=\ci \gamma_0\).
The t-matrix \(t_{n}\) contains no unitarity piece because of an expansion of the denominator according to the power counting: in this channel the unitarity term is included perturbatively at next-to-leading order and beyond~\cite{bedaque03}.

Finally, we discuss briefly how to solve
the stationary Schr\"odinger equation
\begin{equation}
  \Ke{H_0 + \sum_i \K{V_i + V_3^{\K{i}}} }\ket{\Psi} = E_3 \ket{\Psi}
\end{equation}
in the presence of the three-body potentials \(V_3^{\K{i}}\), which are defined with respect to spectator \(i\).
This can be done by adjusting the Faddeev equations.
As explained in \cite{gloeckle83}, one method is to modify Eq. (\ref{eq:def_fd_wfc}) to
\begin{equation}
  \ket{\psi_i} \coloneqq G_0 \K{V_i + V_3^{\K{i}} } \ket{\Psi}\,.
\end{equation}
The Faddeev equations, as defined in Eq. (\ref{eq:fd}), then have to be adjusted.
In practice, the equations that are solved to obtain the \(\ket{F_i}\), \eqref{eq:fd_f}, then change, while the formulas for calculating the \(\ket{\psi_i}\) and
\(\ket{\Psi}\) from the \(\ket{F_i}\) are unaltered.
We use the three-body force as set up in \cite{ji14}. See Ref.~\cite{Ryberg:2017tpv} for alternative possibilities.

To close this section we explain a complication in the calculation of the three-body wave function.
Using the properties of the t-matrices in Eqs. (\ref{eq:t_pw_me}) and (\ref{eq:def_fdt}) we obtain
\begin{align}\label{eq:fdwc_proj_prop}
  \ket{\psi_i} &= \rint{\p} \rint{\q} \sum_{\Omega^\prime} \iket{p,q;\Omega^\prime}{i} \ibraket{i}{p,q;\Omega^\prime}{\psi_i}{} \nonumber \\
  &= \rint{\p} \rint{\q} \iket{p,q;\Omega_i}{i} \ibraket{i}{p,q;\Omega_i}{\psi_i}{} \,.
\end{align}
The Faddeev component $\ket{\psi_i}$ is thus particularly simple in the representation $\iket{p,q;\Omega_i}{i}$.
This simplicity does not extend to the full wave function $\ket{\Psi}$:
 \( \ket{\Psi} \neq \rint{\p} \rint{\q} \iket{p,q;\Omega_i}{i} \ibraket{i}{p,q;\Omega_i}{\Psi}{}\) in general because of the presence of the other two Faddeev components.
When \(\rint{\p} \rint{\q} \sum_{\Omega^\prime} \iket{p,q;\Omega^\prime}{i} \ibraket{i}{p,q;\Omega^\prime}{\psi_j}{}\) is evaluated a recoupling has to be done between different spectators
yielding many non-trivial overlaps.
There is thus no single angular momentum state in the used basis containing the complete angular dependence of the state \(\ket{\Psi}\).
We therefore project on Jacobi momenta plane wave states, and evaluate
the probability density for specific Jacobi momenta as the expectation value of the projection operator \(P_{\vpq} = \ketbra{\vpq}{\vpq}\); this avoids having to choose specific spin states to project onto.
\subsection{Potentials from amplitudes}\label{subsec:coupling_strength}
As already mentioned, we want to calculate the \(\lambda_i\) from the \(\tau_i\); we do so by using Eq.~(\ref{eq:lb_ls}).
The result depends on the chosen form factor. We consider two different functional forms. First:
\begin{equation}
  \reg{{}}{p} = p^l \theta{\K{\beta_l - p}}\,,
\end{equation}
where \(\beta_l\) denotes the regulation scale and \(\theta\) is the Heaviside step function.
We call this the \stdff~here. It is the one used in Ref. \cite{ji14}.
The second is the Yamaguchi form factor
\begin{equation}
  \reg{{}}{p} = p^l \K{1 + \frac{p^2}{\beta_l^2} }^{-\K{l+1}}\,.
\end{equation}
The factor of $p^l$ is required to ensure the t-matrix satisfies the Wigner threshold law~\cite{Wigner:1948zz}.
The rest of the function $\reg{{}}{p}$ determines not only the regularization scheme, but also the off-shell behavior of the t-matrix.
As long as an appropriate three-body interaction is included, observables should not depend on the off-shell behavior of two-body interactions~\cite{Haag:1958vt,Polyzou1990}.
We can test whether this is really the case in our calculation by examining results for different form factors.

In the case of \stdff s~we employ Eqs.~\eqref{eq:sept} and \eqref{eq:lb_ls} in Eqs.~\eqref{eq:t_nn} and \eqref{eq:t_nc} and obtain for the interaction strength:
\begin{align}\label{eq:std_ff_lambda_c}
  \lambda_c{\K{E}} &= \frac{1}{4\pi^2\mu_{nn}} \\
  & \flbc \Ke{\frac{1}{a_0} - \sqrt{-k^2} - \frac{2}{\pi} \beta_0 + \frac{2}{\pi} \sqrt{-k^2} \arccot{\K{\frac{\sqrt{-k^2}}{\beta_0}}} }^{-1}\,, \nonumber \\
  \lambda_n{\K{E}} &= \frac{1}{4\pi^2\mu_{nc}} \label{eq:std_ff_lambda_n} \\
  & \flbc \Ke{ \gamma_1 \K{k^2 - k_R^2} - \frac{2 k^2\beta_1}{\pi } - \frac{2\beta_1^3}{3\pi} - \frac{2}{\pi}\K{-k^2}^{3/2} \arccot{\K{\frac{\sqrt{-k^2}}{\beta_1}}} }^{-1}\,. \nonumber
\end{align}
Note that the expressions depend not only on
\(k^2\) but also on \(\sqrt{-k^2}\).
This is necessary in order to obtain potentials from which the t-matrices are reproduced exactly.
The form of $\lambda_c(E)$ and $\lambda_n(E)$ which is more convenient for bound states is given in Eqs.~\eqref{eq:std_ff_lambda_c} and \eqref{eq:std_ff_lambda_n}.
For scattering problems we would have to analytically continue via \(\sqrt{-k^2} \rightarrow - \mathrm{i} k\).
A unitarity term, \(\mathrm{i}k\), then becomes visible in the equation for \(\lambda_c\), but is canceled by the imaginary part of the analytic continuation of the \(\arccot\).
However, the equation for \(\lambda_n\) does not have a unitarity term outside the \(\arccot\).
This is due to the fact that we use the \(t_{n}\) given in Eq. (\ref{eq:t_nc}), which has no unitarity piece.
$\lambda_c$ and $\lambda_n$ are also different because expanding
the \(\arccot\) in Eqs.~(\ref{eq:std_ff_lambda_c}) and (\ref{eq:std_ff_lambda_n}) in powers of $1/\beta_i$ reveals that, in the limit \(\beta_i \to \infty\),
the energy dependence of \(\lambda_c\) vanishes whereas that of \(\lambda_n\) does not.
This is why a discussion of energy-dependent potentials is unavoidable in that case, see Sec.~\ref{sec:energydeppotentials}.
For the calculations to be carried out there we need derivatives of the \(\lambda_i\) with respect to \(E\).
They are
\begin{align}
 \pdv{\lambda_c}{E} &= -8\pi^2 \mu^2_{nn} \lambda_c^2{\K{E}} \nonumber \\
 & \flbc \Ke{ \frac{1}{2\sqrt{-k^2}}  - \frac{1}{\pi \sqrt{-k^2}} \arccot{\K{\frac{\sqrt{-k^2}}{\beta_0} }}  + \frac{1}{\pi} \frac{\beta_0}{ \beta_0^2 -k^2 } }\,, \\
 \pdv{\lambda_n}{E} &= -8\pi^2 \mu^2_{nc} \lambda_n^2{\K{E}} \nonumber \\
 & \flbc \Ke{ \gamma_1 - \frac{2 \beta_1}{\pi } +\frac{3}{\pi} \sqrt{-k^2}  \arccot{\K{\frac{\sqrt{-k^2}}{\beta_1}}} + \frac{1}{\pi}  \frac{k^2}{\beta_1 - k^2 /\beta_1} } \,.
\end{align}

In the case of Yamaguchi form factors the expressions for the strength parameters are:
\begin{align}
  \lambda_c{\K{E}} &= \frac{1}{4\pi^2\mu_{nn}} \K{ 1 + 2\frac{k^2}{\bn^2} + \frac{k^4}{\bn^4} }
  \Ke{ \frac{1}{a_0} - \frac{\bn}{2} + k^2 \frac{1}{2\bn} }^{-1} 
  \eqqcolon \frac{1}{4\pi^2\mu_{nn} } \frac{u_c}{v_c} \,, \\
  \lambda_n{\K{E}} &= \frac{1}{4\pi^2\mu_{nc}} \K{ 1 + \frac{k^2}{\bo^2} }^4 \nonumber \\
  & \flbc\Ke{ \gamma_1 \K{k^2 -k_R^2} - \frac{\bo^3}{16} \K{ 1 + 9 \frac{k^2}{\bo^2} + 16 \frac{\sqrt{-k^2}^3}{\bo^3} - 9 \frac{k^4}{\bo^4} - \frac{k^6}{\bo^6} } }^{-1}  \nonumber \\
  &\eqqcolon \frac{1}{4\pi^2\mu_{nc} } \frac{u_n}{v_n}\,,
\end{align}
where \(u_c\), \(v_c\), \(u_n\) and \(v_n\) are introduced in order to write the derivatives in a more compact form.
These $\lambda_i$ also reproduce \eqref{eq:t_nc} and \eqref{eq:t_nn} exactly.
Their derivatives are:
\begin{align}
  \pdv{\lambda_c}{E} &= 2 \mu_{nn} \lambda_c \K{ \frac{ \frac{2}{\bn^2} + \frac{2k^2}{\bn^4} }{u_c} - \frac{ \frac{1}{2\bn} }{v_c} }\,, \\
  \pdv{\lambda_n}{E} &= 2 \mu_{nc} \lambda_n \K{ \frac{ 4\K{ 1 + \frac{k^2}{\bo^2} }^3 \frac{1}{\bo^2} }{u_n} - \frac{ - \frac{r_1}{2} - \frac{9}{16}\bo +\frac{3}{2} \sqrt{-k^2} + \frac{9}{8} \frac{k^2}{\bo} + \frac{3}{16} \frac{k^4}{\bo^3} }{v_n} } \,.
\end{align}

In \cite{ji14} similar derivations of the strength parameters were done.
Tuning the regulator parameters \(\beta_0\) and \(\beta_1\) of the Yamaguchi form factors as well as the interaction strengths \(\lambda_c\) and \(\lambda_n\)
Ji {\it et al.} were able to reproduce the effective range parameters \(a_0\), \(r_0\), \(a_1\) and \(r_1\) without energy-dependent
potentials. However, in contrast to what is done here, the inverse t-matrices in \cite{ji14} still contain pieces $\sim 1/\beta_i$. They therefore do not
exactly reproduce the forms \eqref{eq:t_nn} and \eqref{eq:t_nc}, but only do so up to regulator artifacts associated with higher-order terms in the effective-range expansion.
By using energy-dependent potentials we have completely eliminated these artifacts. We have also guaranteed that \(r_1\) is reproduced, no matter what value of the regulator parameter $\beta_1$ is employed.
The implications of this energy dependence are discussed in the following section.
%
\section{Normalization and orthogonality relations for energy-dependent potentials}
\label{sec:energydeppotentials}
\subsection{Orthogonality and normalization relations in the presence of energy-dependent potentials}
In this section we consider the Schr\"odinger Equation for an energy-dependent potential, $V(E)$. We assume that $V$, while energy-dependent, is still Hermitian, i.e.,
\begin{equation}
\langle \phi_1|V(E)|\phi_2 \rangle=\langle \phi_2|V(E)|\phi_1 \rangle^*\,,
\end{equation}
for all real energies $E$ and for all states $|\phi_1 \rangle$ and $|\phi_2 \rangle$.
We define eigenstates of the problem corresponding to the energy $E$ according to:
\begin{equation}
(H_0 + V(E))|\psi_E \rangle=E|\psi_E \rangle\,.
\label{eq:SE}
\end{equation}
with $H_0$ also---as usual---a Hermitian operator. Our discussion of this problem summarizes results from Refs.~\cite{mckellar83,formanek04}.

Consider two eigenstates of $H(E)$ corresponding to different energies, $E_\alpha$ and $E_\beta$ denoted as $|\psi_\alpha \rangle$ and $|\psi_\beta \rangle$, respectively.
Since the two states correspond to the solutions of different eigenvalue problems---one for $H(E_\alpha)$ and one for $H(E_\beta)$---they are not orthogonal.
By projecting the equation for the eigenstate $|\psi_\alpha \rangle$ onto $|\psi_\beta \rangle$, and vice versa, and taking the difference of the results, we find:
\begin{equation}
\langle \psi_\beta|V(E_\alpha) - V(E_\beta)|\psi_\alpha \rangle=(E_\alpha - E_\beta) \langle \psi_\beta|\psi_\alpha \rangle\,,
\end{equation}
which may be rewritten as:
\begin{equation}
\langle \psi_\beta|(\id - \Delta V_{\beta \alpha})|\psi_\alpha \rangle=0\,,
\label{eq:orthogonality}
\end{equation}
with
\begin{equation}
\Delta V_{\beta \alpha} \coloneqq \frac{V(E_\alpha) - V(E_\beta)}{E_\alpha - E_\beta}\,.
\end{equation}
Obviously $\Delta V=0$ for an energy-independent potential. But, in a problem with an energy-dependent potential, Eq.~(\ref{eq:orthogonality}) states that
the orthogonality relation is only obtained if, to the usual ``1" that is inserted between the two states, we add the contribution from $\Delta V$.
Alternatively, we can say that $\langle \psi_\beta|$ is orthogonal to $(1 - \Delta V_{\beta \alpha})|\psi_\alpha \rangle$ and/or $|\psi_\alpha \rangle$ is orthogonal to $\langle \psi_\beta|(1 - \Delta V_{\beta \alpha})$.
This is related to the modifications to the standard scalar product for states bound by a pseudopotential discussed by Pricoupenko in Refs.~\cite{Pricoupenko:2006A,Pricoupenko:2006B}.

We may use this interpretation as the starting point for construction of the operator $D^{-1}$, introduced by McKellar and McKay~\cite{mckellar83}, that converts the set of states $\langle \psi_\beta|$ into a set of biorthogonal states. I.e., we seek $D$ such that:
\begin{equation}
\langle \langle \psi_\beta|=\langle \psi_\beta|D^{-1}
\end{equation}
with
\begin{equation}
\langle \langle \psi_\beta|\psi_\alpha \rangle=\delta_{\beta \alpha}\,.
\end{equation}
It is then tempting to identify $D^{-1}=1- \Delta V_{\beta \alpha}$.
However---as pointed out by Form\'anek {\it et al.}~\cite{formanek04}---this identification is not correct in general, because $\Delta V_{\beta \alpha}$ is an energy-dependent, and hence-state dependent, operator.  McKellar and McKay show how to construct $D^{-1}$ as the solution to an integral equation.

With the operator $D^{-1}$ in hand we can write a completeness relation:
\begin{equation}
\id=\sum_\alpha |\psi_\alpha \rangle \langle \langle \psi_\alpha|=\sum_\alpha |\psi_\alpha \rangle \langle \psi_\alpha| D^{-1}\,.
\label{eq:id}
\end{equation}
The presence of $D^{-1}$ here has an important implication: the set of states $\{|\psi_\alpha \rangle\}$, while complete, is not orthornormal, and so {\it cannot} be inserted as an identity when deriving quantum-mechanical equations.
This is crucial for some forms of scattering theory~\cite{mckellar83}, but in the derivations performed in Sec.~\ref{sec:6HeHaloEFTreview} only complete sets of plane waves were inserted, so it does not affect any of the results presented above.

It does, though, affect the normalization condition for the three-body wave function, and hence the definition of the momentum density.
From the resolution of the identity operator, (\ref{eq:id}), we  identify the projector onto the eigenstate of $H(E)$ corresponding to energy $E_\alpha$ as
\begin{equation}
P_\alpha=|\psi_\alpha \rangle \langle \langle \psi_\alpha|=|\psi_\alpha \rangle \langle \psi_\alpha| D^{-1}\,.
\label{eq:projector}
\end{equation}
For $P_\alpha$ defined according to Eq.~(\ref{eq:projector}) to be a standard projector onto a subspace of the Hilbert space of eigenstates of $H(E)$ we need $P_\alpha^2=P_\alpha$. This will only be true if:
\begin{equation}
\langle \psi_\alpha| D^{-1}|\psi_\alpha \rangle=1.
\end{equation}

In the case of a linear-in-energy potential, $V(E)=V_0 + V_1 E$, the argument becomes much  simpler, since then $\Delta V_{\beta \alpha}=V_1$ is state independent, and $D^{-1}=\id-V_1$ at the operator level. Equation (\ref{eq:projector}) then reduces to:
\begin{equation}
P_\alpha=|\psi_\alpha \rangle \langle \psi_\alpha| (\id - V_1)\,,
\label{eq:Palpha}
\end{equation}
and the normalization condition is:
\begin{equation}
\langle \psi_\alpha|(\id - V_1)|\psi_\alpha \rangle=1\,.
\label{eq:normconditionlinear}
\end{equation}

This result permits us to construct the diagonal matrix elements of $D^{-1}$ for an arbitrary $V(E)$. For energies $E \approx E_\alpha$ we can replace $V(E) \approx V(E_\alpha) + V'(E_\alpha) (E-E_\alpha)$. Following the steps of the previous paragraph then gives the normalization of the state $|\psi_\alpha \rangle$ as:
\begin{equation}
\langle \psi_\alpha|(\id - V'(E_\alpha))|\psi_\alpha \rangle=1\,.
\label{eq:normcondition}
\end{equation}
Form\'anek {\it et al.} arrive at the same conclusion by considering the continuity equation in the presence of an energy-dependent potential~\cite{formanek04}. For yet another derivation, see Appendix~\ref{ap:alternativederivationnorm}.
From now on we refer to the normalization condition $\langle \psi_\alpha|\psi_\alpha \rangle=1$ as the ``naive" normalization condition and Eq.~(\ref{eq:normcondition}) as the ``correct" normalization.

The projector $P_\alpha$ is not a Hermitian operator. So simply evaluating diagonal matrix elements of $P_\alpha$ using an eigenstate of the momentum operator does not lead to a momentum-space probability density with expected properties, e.g.,
$\rho_\alpha({\bf p}) \geq 0$. As pointed out by Form\'anek {\it et al.}, if it is to correspond to an observable, the operator $O$ should be Hermitian with respect to matrix elements $\langle \langle \phi|O|\psi \rangle$. This means that $O$ must equal $O^\#$ with the operator $O^\#$ defined by its matrix elements as:
\begin{equation}
\langle \langle \psi| O^\# |\phi \rangle \coloneqq \langle \langle \phi|O|\psi \rangle^*\,,
\end{equation}
or, in operator form:
\begin{equation}
O^\# \coloneqq D O^\dagger D^{-1}\,,
\end{equation}
with $^\dagger$ the usual Hermitian conjugation. (Here we have used, without proof, that $D=D^\dagger$.)
The simplest extension of the usual momentum-space projector $|{\bf p} \rangle \langle {\bf p}|$ which obeys this requirement is
\begin{equation}
P_{\bf p}=\sqrt{D} |{\bf p} \rangle \langle {\bf p}| \sqrt{D^{-1}}\,.
\label{eq:PpD}
\end{equation}
Note that ${\rm Tr}(P_{\bf p})=1$ for a set of normalized momentum-space eigenstates. Note also that Eq.~(\ref{eq:PpD}) is only well defined if $D^{-1}$ is a positive semi-definite operator.

We now construct the momentum-space probability density by taking the trace of the product of the momentum-space projector (\ref{eq:PpD}) and the bound-state projector (\ref{eq:Palpha}),
which acts as a pure-state density matrix.
In general this gives
\begin{equation}
\rho_{\alpha}({\bf p})=\langle \psi_\alpha|\sqrt{D^{-1}}|{\bf p} \rangle \langle {\bf p}|\sqrt{D^{-1}}|\psi_\alpha \rangle=|\langle {\bf p}|\sqrt{D^{-1}}|\psi_\alpha \rangle|^2\,,
\label{eq:exactpd}
\end{equation}
which is non-negative, as expected\footnote{
  This result for the probability density can also be derived without invoking a density matrix
  if we instead compute the expectation value in the ${}^6$He ground state of the modified momentum-space projector (\ref{eq:PpD}) using the modified scalar product of \cite{formanek04}.
}.
Constructing the operator $\sqrt{D^{-1}}$ is, in general, somewhat complicated for the three-body problem.
In what follows we will exploit the fact that, for the momenta at which the EFT is valid and for cutoffs $\Lambda$ well above the EFT breakdown scale, $V'(E_\alpha) \ll 1$ (see explicit formulas in Sec.~\ref{sec:6HeHaloEFTreview}).
In that case the formal expression involving $\sqrt{D^{-1}}$ can be replaced by
\begin{equation}
\rho_{\alpha}({\bf p})=|\langle {\bf p}|\sqrt{\id-V'(E_\alpha)}|\psi_\alpha \rangle|^2\,.
\label{eq:density}
\end{equation}
This density is positive semi-definite and is also normalized to 1, provided that $|\psi_\alpha \rangle$ is normalized according to Eq.~(\ref{eq:normcondition}).
But in the regime where $V'(E_\alpha) \ll 1$ the density (\ref{eq:density}) can be further approximated as:
\begin{equation}
\rho_{\alpha}({\bf p})\approx |\langle {\bf p}|\psi_\alpha \rangle|^2 - \Re(\langle {\bf p} |V'(E_\alpha)|\psi_\alpha \rangle \langle \psi_\alpha|{\bf p} \rangle)\,.
\label{eq:approxdensity}
\end{equation}
Note that:
\begin{enumerate}
\item If such an evaluation of $\rho_\alpha$ yields a negative number it signifies that we are outside the domain of validity of the expansion used to justify it.

\item The approximation leading to (\ref{eq:approxdensity}) preserves the normalization of $\rho_\alpha$, thanks to the normalization condition (\ref{eq:normcondition}).
\end{enumerate}

The above derivation considered eigenstates of a single momentum operator.
In the three-body problem we construct the momentum-space projector
for a specific Jacobi representation and get:
\begin{equation}
\rho_i(\vpq ) \approx \ibraket{i}{ \vpq }{\Psi}{} \left( \ibraket{}{ \Psi}{ \vpq }{i}  - \Re\left[ \sum_j \imel{}{ \Psi \vphantom{\frac{q^2}{2\mu_{i\K{jk}}}} }{ V_j'{\K{E_\alpha- \frac{q_j^2}{2\mu_{j\K{ki}}} }} }{ \vpq }{i} \right]\right)\,,
\label{eq:Threebodydensity}
\end{equation}
where the subscript $i$ now indicates the Jacobi momenta ${\bf p}$ and ${\bf q}$ for which we are computing $\rho$, and {\it not} the state of interest.
We have assumed that the three-body potential, $V_3$, is energy independent, but have not assumed anything about which two-body subsystem(s) the energy-dependent potential is active in. If $j \neq i$ recoupling is necessary in order to compute the matrix element of $V_j$ here.

Finally we reiterate that, while the naive probability density contains only the one-body part $|\psi|^2$ (or $|\Psi|^2$), our corrected results, Eqs.~(\ref{eq:approxdensity}) and (\ref{eq:Threebodydensity}), include two-body contributions.
\subsection{Example: two-body bound state with an energy-dependent contact interaction}\label{subsec:ve_tb_example}
The EFT for a two-body system in which both the scattering length and effective range are unnaturally large~\cite{beane00} can be implemented by considering an energy-dependent potential whose matrix elements are:
\begin{equation}
\langle \v{p}|V(E)|\v{p}' \rangle=\frac{1}{\Delta_0 + \Delta_2 E}\,,
\end{equation}
with $\Delta_0$ and $\Delta_2$ parameters that, as we will see below, are related to the bound-state pole position and residue.
If $V$ is interpreted as due to an $s$-channel particle exchange then $\Delta_0$ is proportional to the particle's mass shift, and $\Delta_2$ is related to its kinetic mass~\cite{kaplan97}.
Reference \cite{beane00} showed that the charge density for such a system includes both one- and two-body terms, but is then properly normalized. Here we demonstrate that the normalization condition (\ref{eq:normcondition}) has the same property.

First we compute the two-body t-matrix for this problem: $t(E)=V(E) + V(E) G_0(E) t(E)$. Since $V(E)$ is independent of momentum $t(E)$ is too, and the Lippmann-Schwinger equation becomes algebraic.
The solution is:
\begin{equation}
\langle \v{p}|t(E)|\v{p}'\rangle=\frac{1}{\Delta_0 + \Delta_2 E + 8 \pi m_R \Lambda + \ci 4 \pi^2 m_R k}\,,
\end{equation}
with $m_R$ the two-body reduced mass, $k=\sqrt{2 m_R E}$ the on-shell momentum, and $\Lambda$ a sharp cutoff used to make the integral of $G_0(E)$ finite. We adjust $\Delta_0$ so that the system has one bound state, at $E=-B$. $t$ then takes the form:
\begin{equation}
\langle \v{p}|t(E)|\v{p}' \rangle=\frac{1}{4 \pi^2 m_R} \frac{1}{1 - \gamma r}\frac{1}{\gamma + \ci k}\,,
\end{equation}
with:
\begin{equation}
\frac{\Delta_0}{4 \pi^2 m_R}=\gamma - \frac{2\Lambda}{\pi} - \frac{r \gamma^2}{2}; \qquad \frac{\Delta_2}{4 \pi^2 m_R}=-r m_R \,.
\end{equation}
where $r$ is the effective range for the expansion around the deuteron pole.
Note that the second equation has pieces of ${\cal O}(1/\Lambda)$ if we include the regulator in the potential.
In the calculation of the coupling strengths in subsection \ref{subsec:coupling_strength} this effect of the $s$-wave regulator (there denoted $\beta_0$) was taken into account.

The three-dimensional wave function $\psi$ obtained directly from $t(E)$, without imposing any normalization, is:~\cite{phillips02,hammer17}
\begin{equation}
\langle {\bf p}|\psi \rangle=\frac{1}{\pi} \frac{\sqrt{{\cal Z} \gamma}}{p^2 + \gamma^2}
\end{equation}
where we have defined ${\cal Z}=1/(1 - \gamma r)$ as the wave function renormalization, because
\begin{equation}
\int \dd^3p |\langle {\bf p}|\psi \rangle|^2={\cal Z}\,.
\end{equation}
So the wave function is not normalized according to the naive normalization condition.

Now
\begin{equation}
\langle {\bf p}'| V'(-B)|{\bf p}\rangle=-\frac{\Delta_2}{(\Delta_0 - \Delta_2 B)^2}= \frac{r}{4 \pi^2} \frac{1}{(\gamma - \frac{2 \Lambda}{\pi})^2}
\end{equation}
Therefore the two-body momentum density in the region of validity of the EFT, Eq.~(\ref{eq:approxdensity}) is approximately:
\begin{equation}
\rho({\bf p}) \approx \frac{{\cal Z} \gamma}{\pi^2}\frac{1}{p^2 + \gamma^2} \left[\frac{1}{p^2 + \gamma^2} - \frac{r}{2} \frac{1}{(\gamma - \frac{2 \Lambda}{\pi})}\right]\,,
\label{eq:rhop}
\end{equation}
where we have regulated the divergent integral encountered in the two-body contribution to the probability density using
the same sharp cutoff as was employed in the calculation of the t-matrix. (In fact, the same divergent integral
that appears everywhere in this analysis, $\int \frac{\dd^3p}{\sqrt{(2 \pi)^3}} \psi({\bf p})=\psi({\bf r}=0)$.)
Integrating Eq.~(\ref{eq:rhop}) over all momenta (and again regulating the second term using a sharp cutoff) we find:
\begin{equation}
\int \dd^3p \rho({\bf p})={\cal Z}[1 - r \gamma]=1\,.
\end{equation}
This verifies that the normalization condition (\ref{eq:normcondition}) produces a properly normalized probability density for this problem.

To close this section we note a few points that are relevant for our application of this formalism to ${}^6$He in the next section
\begin{itemize}
\item For momenta $p \sim \gamma$ the two-body contribution to $\rho({\bf p})$ is a correction to the one-body piece of ${\cal O}\left(\frac{r \gamma^2}{\Lambda}\right)$.
For large cutoffs it is thus negligible at low momentum. This justifies the linear approximation used to evaluate $\rho({\bf p})$.

\item For $p \sim \Lambda$ the two-body contribution is larger than the one-body contribution by a factor $r \Lambda$.

\item In the approximate form of $\rho$ the two-body part of the momentum-space probability density is negative. This is to be expected since ${\cal Z} > 1$.
But, since the positive one-body piece dominates at small momentum, it follows that $\rho({\bf p})$ will go through zero at a momentum $p_0 \sim \sqrt{\Lambda/r}$.
For sufficiently large $\Lambda$ this is well above the breakdown scale of the EFT; the expression we are using for $\rho({\bf p})$ is not reliable for $p \sim p_0$.

\item The two-body contribution to the norm is proportional to a divergent integral. This, combined with the ${\cal O}(1/\Lambda)$ coefficient, produces a finite effect.
\end{itemize}
%
\section{Results for the probability density}
\label{sec:probdensity}
\subsection{Calculation of one-body probability density}\label{subsec:naive_pbd}
The probability density is given by
\begin{align}\label{eq:prob_dens}
  \rhos{i}{\vpq} &\coloneqq \mel{\Psi}{\K{ \iketbra{\vpq}{i}{} \otimes \spid }}{\Psi} \quad \textrm{or}\\
  \rhot{i}{\vpq} &\coloneqq \mel{\Psi}{\K{ \iketbra{\vpq}{i}{} \otimes \spid }\K{\id - \pdv{V}{E_3}}}{\Psi}\label{eq:prob_dens_mn}
\end{align}
in the presence of an energy-dependent potential.
Note that, in order to define the projection operator, we include an identity in spin space.
Hence, the probability density we compute here is averaged over all allowed spin states.
\(\ket{\Psi}\) is calculated from \(\ket{F_i}\) via the equations \(\ket{\Psi} = \ket{\psi_n} + \ket{\psi_{n^\prime}} + \ket{\psi_c}\) and \(\ket{\psi_i} = G_0 t_i \ket{F_i}\).
Since we obtained \(F_i{\K{q}} \coloneqq \rint{\p} \reg{i}{p} \ibraket{i}{p,q;\Omega_i}{F_i}{}\)\,, it is useful to insert an identity between the t-matrix and
the Faddeev amplitude in the partial wave representation.
Due to the projection property of the t-matrix it also useful to insert one identity on its left.
We then obtain
\begin{align}\label{eq:psi_i_pw}
    \sum_{\Omega} \iket{p,q;\Omega}{i} \ibraket{i}{p,q;\Omega}{\psi_i}{} &= \iket{p,q;\Omega_i}{i} 4\pi \gz{i}{p,q;-B_3^{\K{0}}} \reg{i}{p} \rtm{i}{q}{-B_3^{\K{0}}} \nonumber \\
    &\flbc \rint{\pp} \reg{i}{\pp} \ibraket{i}{\pp,q;\Omega_i}{F_i}{} \nonumber \\
    &= \iket{p,q;\Omega_i}{i} 4\pi \gz{i}{p,q;-B_3^{\K{0}}} \reg{i}{p} \rtm{i}{q}{-B_3^{\K{0}}} F_i{\K{q}} \nonumber \\
    &= \iket{p,q;\Omega_i}{i} \psi_i{\K{p,q}} \,,
\end{align}
where the definition \(\rtm{i}{q}{E_3} \coloneqq \tau_i{\K{E_3-q^2/\K{2\mu_{i\K{jk}}}}}\) is used and
the component wave function \(\psi_i{\K{p,q}} \coloneqq \ibraket{i}{p,q;\Omega_i}{\psi_i}{}\) is introduced.
In order to evaluate the expressions for the probability density given in Eq. (\ref{eq:prob_dens}) or in Eq. (\ref{eq:prob_dens_mn}) using this expression
one has to decouple \(\ket{\Omega_i}_i\) into the angular and spin part as given in \cite{ji14}.
The emerging overlaps of plane wave states with partial wave states are given by coupled spherical harmonics
\begin{align}
    \cy{l \lambda}{LM}{\v{\hat{p}}, \v{\hat{q}}} &\coloneqq \braket{\vpq}{p,q;\K{l \lambda} LM}\\
    &=\sum_{\mathclap{m_l+m_\lambda=M}} \cgc{l m_l \lambda m_\lambda}{L M} \y{l}{m_l}{\v{\hat{p}}} \y{\lambda}{m_\lambda}{\v{\hat{q}}}\,,
\end{align}
where \(\y{l}{m_l}{\v{\hat{p}}}\) is the usual spherical harmonic.
Using Eq. (\ref{eq:psi_i_pw}) we obtain\footnote{
  Note that we abbreviate the spin states in the following way:
  \begin{align*}
    \iket{L, -M}{n} &= \iket{\K{\oh, 0}\oh,\oh; L, -M}{n}\,, \\
    \iket{0, 0}{c} &= \iket{\K{\oh, \oh}0,0; 0, 0}{c}\,,
  \end{align*}
  where the notation \(\iket{\K{\nu_j,\nu_k}s_i, \sigma_i; S, M_S}{i}\) from \cite{ji14} is used.
}
\begin{align}\label{eq:ov_plw_psi}
    \ibraket{i}{\vpq}{\Psi}{} &= \sum_{\Omega} \ibraket{i}{\vpq}{p,q;\Omega}{i} \ibraket{i}{p,q;\Omega}{\Psi}{} \nonumber \\
    &= a_i \spiso{L}{M} c_{L,M}\cy{11}{LM}{\ya_i} \ket{L,-M}_n \nonumber \\
    &\flb- \tilde{a}_i \spiso{L}{M} c_{L,M}\cy{11}{LM}{\tilde{\ya}_i} \pmospin \ket{L,-M}_n + \frac{d_i}{4\pi} \ket{0,0}_c\,,
\end{align}
where \(\pmospin\) is the spin space part of the \(nn\) commutation operator \(\pmo\).
The coefficients \(c_{L,M}\) result from the recoupling from \jJcoupling~to \LScoupling~given in \cite{ji14}.
Thus they do not depend on the spectator.
They are given by \(c_{L,M} \coloneqq \K{-1}^{M} \sqrt{\frac{2^{1-L}}{6L+3}}\) with \(L \in \{0,1\}\) and \(-L \leq M \leq L\).
Note that the result of Eq. (\ref{eq:ov_plw_psi}) is the sum of scalars times spin states.
The other parameters, which depend on the spectator \(i\), are given by
\begin{align}\label{eq:pbd_coeff_n_first}
  a_n &= \psi_n{\K{p,q}}\,, \\
  \tilde{a}_n &= \psi_n{\K{\knnpp,\knnqp}}\,, \\
  d_n &=  \psi_c{\K{\kncp, \kncq}}\,, \\
  \ya_n &= \K{\v{\hat{p}}, \v{\hat{q}}}\,, \\
  \tilde{\ya}_n &= \K{ \knnpph, \knnqph }\,.
\end{align}
for the \(n\) as spectator.
The definition of the \(\kappa\) are given in appendix \ref{subsec:jacobi}.
The functions in \(a_i\) depend on the norm of the vectors which are contained in \(\ya_i\), the analogous relation holds for \(\tilde{a}_i\) and \(\tilde{\ya}_i\).
Subsequently the expressions for \(c\) as spectator are given:
\begin{align}
  a_c &= \psi_n{\K{\kcnp, \kcnq}}\,, \\
  \tilde{a}_c &= \psi_n{\K{\kcnp^\prime, \kcnq^\prime}}\,, \\
  d_c &= \psi_c{\K{p,q}}\,, \\
  \ya_c &= \K{\kcnph, \kcnqh}\,, \\
  \tilde{\ya}_c &= \K{\kcnpph, \kcnqph}\,. \label{eq:pbd_coeff_c_last}
\end{align}
Based on Eq. (\ref{eq:ov_plw_psi}) and the following equations we obtain the expression below for the one-body probability density by using the overlaps of spin states.
Most of these overlaps are given in \cite{ji14}.
\begin{align}\label{eq:ob_pbd}
    \rho_i{\K{\v{p},\v{q}}} &= \mel{\Psi}{\K{\iketbra{\vpq}{i}{} \otimes \spid }}{\Psi} \nonumber \\
    &= \bigg[ \spiso{L}{M} c_{L,M}^2 \K{ a_i^2 \left| \cy{11}{LM}{\ya_i} \right|^2 + \tilde{a}_i^2 \left| \cy{11}{LM}{\tilde{\ya}_i} \right|^2 } + \K{\frac{d_i}{4\pi}}^2  \nonumber \\
    & \flb - 2a_i \tilde{a}_i \spiso{L}{M} c_{L,M}^2 \K{-1}^{1-L} \Re\K{\cy{11}{LM}{\tilde{\ya}_i}^* \cy{11}{LM}{\ya_i}} \nonumber \\
    & \flb - 2\tilde{a}_i \frac{d_i}{4\pi} c_{0,0} \cy{11}{00}{\tilde{\ya}_i} -2 a_i \frac{d_i}{4\pi} c_{0,0} \cy{11}{00}{\ya_i}   \bigg]\,.
\end{align}
We used that \(a_i\), \(\tilde{a}_i\) and \(d_i\) are real.

Because \(c_{L,M}^2\) is independent of \(M\)
it follows that \(\rho_i{\K{\vpq}}\) only depends on \(p\), \(q\) and \(\theta_{pq} \coloneqq \arccos{ \K{\v{p} \cdot \v{q}/\K{pq}}} \).
The simplification is very helpful, since it reduces the computational effort of angular integrals of the probability density.
Additional remarks on it can be found in appendix \ref{subsec:ang_simp}.
It also motivates an expansion in \(\cos{\theta_{pq}}\):
\begin{align}
  \rhopw{i}{\xi}{p, q} &= \int_{0}^\pi \dd{\theta_{pq}} \sin{\theta_{pq}} P_{\xi}{\K{\cos{\theta_{pq}}}} \rhos{i}{\vpq}\,, \label{eq:rhopw} \\
  \rhotpw{i}{\xi}{p, q} &= \int_{0}^\pi \dd{\theta_{pq}} \sin{\theta_{pq}} P_{\xi}{\K{\cos{\theta_{pq}}}} \rhot{i}{\vpq}\,, \label{eq:rhotpw}
\end{align}
where \(P_\xi\) denotes the \(\xi\)-th Legendre polynomial.
Although \(\xi\) looks like an
angular-momentum quantum number there is, in fact, no wave function component angular momentum---neither \(l\) nor \(\lambda\) nor \(L\)---that can be associated with it.
\subsection{Calculation of two-body probability density}\label{subsec:mod_pbd}
We proceed by calculating an approximation to the two-body probability density as given in Eq. (\ref{eq:prob_dens_mn}), which accounts for the energy dependence of the potentials.
For simplicity we call it just the two-body probability density and do not stress its approximate character.
In subsection \ref{subsec:coupling_strength} the expressions for the coupling strengths of the potentials \(V_c\) and \(V_n\) were already derived.
As explained in subsection \ref{subsec:heft}, for \(V_{\np}\) the equation
\begin{align}
  V_{\np} = \K{- \pmo} V_n \K{- \pmo}
\end{align}
holds. In following calculations we will use \(V_{\np}\ket{\Psi} = \K{- \pmo} V_n \ket{\Psi}\).

Since \(V = V_n + V_{\np} + V_c + V^{\K{3}}\) and the three-body potential \(V^{\K{3}}\) is energy indepedendent, the derivative of the potential with respect to the energy
is given by
\begin{equation}
  \pdv{V}{E_3} = \pdv{V_n}{E_3} + \pdv{V_{\np}}{E_3} + \pdv{V_c}{E_3}\,.
\end{equation}
It is useful to split the calculation in the following way:
\begin{align}\label{eq:prob_dens_mn_dt}
  &\mel{\Psi}{\K{ \iketbra{\vpq}{i}{} }\K{\id - \pdv{V}{E_3}}}{\Psi} = \mel{\Psi}{\K{ \iketbra{\vpq}{i}{} }}{\Psi}  \nonumber \\
  & \flb - \mel{\Psi}{\K{ \iketbra{\vpq}{i}{} }\pdv{V_n}{E_3}}{\Psi} - \mel{\Psi}{\K{ \iketbra{\vpq}{i}{} }\pdv{V_{\np}}{E_3}}{\Psi} \nonumber \\
  & \flb - \mel{\Psi}{\K{ \iketbra{\vpq}{i}{} }\pdv{V_c}{E_3}}{\Psi}\,,
\end{align}
where \(\iketbra{\vpq}{i}{} \otimes \spid\) is abbreviated by \(\iketbra{\vpq}{i}{}\).
The first term is the one-body probability density we already computed.
In the case of the other terms we can generalize the expression we have to evaluate a bit.
We consider the calculation of
\begin{equation}\label{eq:def_O_i_dens}
  O_i{\K{\vpq}} \coloneqq \mel{\Psi}{\K{ \iketbra{\vpq}{i}{} \otimes \spid }O_i}{\Psi}\,,
\end{equation}
where \(O_i\) can be either a potential or its derivative.
Only the properties these have in common are used in further simplifications:
\begin{equation}
  \imel{i}{p,q;\Omega}{O_i}{\pp,\qp;\Omega^\prime}{i} \propto \kd{\Omega}{\Omega^\prime} \kd{\Omega}{\Omega_i} \frac{\de{q-\qp}}{q \qp} \,.
\end{equation}
This enables us the reuse the result for the calculation of ``potential energy densities"\footnote{
  We put it in quotation marks, since it is actually only an approximation for the potential energy density.
  The potential energy density of \(V_i\) is defined as \(P_{\vpq} V_i\),
  where \(P_{\vpq}\) projects on the Jacobi momenta, \(\v{p}\) and \(\v{q}\).
  The formula above represents the expectation value of this product of operators
  in the case of potentials which are not energy dependent.
  Otherwise, correction terms exist.
}
\(\mel{\Psi}{\K{ \iketbra{\vpq}{i}{} }{V_i}}{\Psi}\).
Nevertheless, we want to give on this occasion the full matrix elements of the two-body potentials \(V_i\) when
embedded in a three-body system:
\begin{align}
  \imel{i}{p,q;\Omega}{V_i{\K{E_3}}}{\pp,\qp;\Omega^\prime}{i} &= 4\pi \reg{i}{p} \lambda_i{\left(E_3 - \frac{q^2}{2 \mu_{i\K{jk}}}\right)} \reg{i}{\pp} \nonumber \\
    & \flbc \kd{\Omega}{\Omega^\prime} \kd{\Omega}{\Omega_i} \frac{\de{q-\qp}}{q^2}\,.
\end{align}
This expression also directly determines the matrix elements of \(V_i^\prime\).

Our expressions for the different densities will be expressed using the formula
\begin{align}
  O_i{\K{\vpq}} &\coloneqq \mel{\Psi}{\K{ \iketbra{\vpq}{i}{} \otimes \spid }O_i}{\Psi} \nonumber \\
  &= \bigg[ \spiso{L}{M} c_{L,M}^2 \K{ a_i a_i^\prime \left| \cy{11}{LM}{\ya_i} \right|^2 + \tilde{a}_i \tilde{a}_i^\prime \left| \cy{11}{LM}{\tilde{\ya}_i} \right|^2 } + \frac{d_i d_i^\prime}{\K{4\pi}^2} \nonumber \\
  & \flb - \K{a_i \tilde{a}_i^\prime + \tilde{a}_i a_i^\prime} \spiso{L}{M} c_{L,M}^2 \K{-1}^{1-L} \Re\K{\cy{11}{LM}{\tilde{\ya}_i}^* \cy{11}{LM}{\ya_i}} \nonumber \\
  & \flb - \K{\tilde{a}_i d_i^\prime + d_i \tilde{a}_i^\prime} \frac{1}{4\pi} c_{0,0} \cy{11}{00}{\tilde{\ya}_i} -\K{a_i d_i^\prime + d_i a_i^\prime} \frac{1}{4\pi} c_{0,0} \cy{11}{00}{\ya_i}   \bigg]\,.
\end{align}
We used that \(a_i\), \(a_i^\prime\), \(\tilde{a}_i\), \(\tilde{a}_i^\prime\), \(d_i\) and \(d_i^\prime\) are real.
For the coefficients \(a_i\), \(\tilde{a}_i\) and \(d_i\) as well as for \(\ya_i\) and \(\tilde{\ya}_i\) Eqs. (\ref{eq:pbd_coeff_n_first}) to (\ref{eq:pbd_coeff_c_last}) hold.
The coefficients \(a^\prime_i\), \(\tilde{a}^\prime_i\) and \(d^\prime_i\) involve matrix elements of \(O_i\).
In order to evaluate \(\imel{i}{p,q;\Omega_i}{O_i}{\psi_j}{}\) in the case \(j \neq i\) a recoupling in the partial wave basis has to be made.
Due to the properties of \(\ket{\psi_j}\) the relation
\begin{align}
  \imel{i}{p,q;\Omega_i}{O_i}{\psi_j}{} &= \rint{\pp} \imel{i}{p,q;\Omega_i}{O_i}{\pp,q;\Omega_i}{i} \nonumber \\
  & \flb \cross \rint{\ptp} \rint{\qtp} \ibraket{i}{\pp,q;\Omega_i}{\ptp,\qtp;\Omega_j}{j}
  \ibraket{j}{\ptp,\qtp;\Omega_j}{\psi_j}{}
\end{align}
holds.
Accordingly overlaps of partial wave states with different spectators have to be evaluated.
Helpful identities are given in appendix \ref{subsec:ov_pwb}.
With these the following relations for \(O_n\) can be derived:
\begin{align}
    a_n^\prime &= \rint{\pp} \imel{n}{p,q;\Omega_n}{O_n}{\pp,q;\Omega_n}{n} \psi_n{\K{\pp,q}} \nonumber \\
    & \flb - \rint{\pp} \imel{n}{p,q;\Omega_n}{O_n }{\pp,q;\Omega_n}{n} \nonumber \\
    & \flb \flb \cross \angint{\pp} \angint{\qt} \sum_{L=0}^1 \sum_{M=-L}^{L} \K{-1}^{L} \frac{2^{1-L}}{6L+3} \nonumber \\
    & \flb \flb \cross \K{\cy{11}{LM}{\v{\pp},\v{\qt}}}^* \cy{11}{LM}{\knnppv,\knnqpv} \psi_n{\K{\knnpp,\knnqp}} \nonumber \\
    & \flb + \frac{ 1  }{\sqrt{2}} \rint{\pp} \imel{n}{p,q;\Omega_n}{O_n}{\pp,q;\Omega_n}{n} \nonumber \\
    & \flb \flb \cross \int_{0}^{\pi} \dd{\theta_{\v{\pp},\v{\qt}}}  \sin{\theta_{\v{\pp},\v{\qt}}} \cos{\theta_{\v{\pp},\v{\qt}}} \psi_c{\K{\kncp,\kncq}}\,. \\
    \tilde{a}_n^\prime &= 0\,, \\
    d_n^\prime &= 0\,.
\end{align}
Note that the arguments of the functions \(\kappa_{ijk}\) are omitted.
They are functions of the momenta \(\v{\pp}\) and \(\v{\qt}\).
The vector \(\v{\qt}\) has the same absolute value as \(\v{q}\), it is introduced since
\(\v{q}\) is already in use.
For the angular integration \(\angint{\pp} \angint{\qt}\) in the expression above the identities given
in appendix \ref{subsec:ang_simp} can be used, so that only one angular integral has to be carried out numerically.
In order to calculate \(O_c\) one needs:
\begin{align}
    a_c^\prime &= 0\,, \\
    \tilde{a}_c^\prime &= 0\,, \\
    d_c^\prime &= \rint{\pp} \imel{c}{p,q;\Omega_c}{O_c}{\pp,q;\Omega_c}{c} \psi_c{\K{\pp,q}} \nonumber \\
    & \flb + \frac{ 1  }{\sqrt{2}} \rint{\pp} \imel{c}{p,q;\Omega_c}{O_c}{\pp,q;\Omega_c}{c} \nonumber \\
    & \flb \flb \cross \int_{-1}^{1} \dd{ \cos{\theta_{\v{\pp},\v{\qt}}} }  \cos{\theta_{\kcnpv, \kcnqv}} \psi_n{\K{\kcnp,\kcnq}} \nonumber \\
    & \flb +  \frac{ 1  }{\sqrt{2}} \rint{\pp} \imel{c}{p,q;\Omega_c}{O_c}{\pp,q;\Omega_c}{c} \nonumber \\
    & \flb \flb \cross \int_{-1}^{1} \dd{ \cos{\theta_{\v{\pp},\v{\qt}}} }  \cos{\theta_{\kcnppv, \kcnqpv}} \psi_n{\K{\kcnpp,\kcnqp}}
\end{align}
Also in this case \(\kappa_{ijk}\) are functions of \(\v{\pp}\) and \(\v{\qt}\).
Again, \(|\v{\qt}| = |\v{q}|\) holds.
The density \(O_{\np}\) can be obtained with
\begin{align}
  a^\prime_{\np} &= 0\,, \\
  \tilde{a}^\prime_{\np} &= \Bigg( \rint{\pp} \imel{n}{p,q;\Omega_n}{O_n}{\pp,q;\Omega_n}{n} \psi_n{\K{\pp,q}} \nonumber \\
  & \flb - \rint{\pp} \imel{n}{p,q;\Omega_n}{O_n }{\pp,q;\Omega_n}{n} \nonumber \\
  & \flb \flb \cross \angint{\pp} \angint{\qt} \sum_{L=0}^1 \sum_{M=-L}^{L} \K{-1}^{L} \frac{2^{1-L}}{6L+3} \nonumber \\
  & \flb \flb \cross \K{\cy{11}{LM}{\v{\pp},\v{\qt}}}^* \cy{11}{LM}{\knnppv,\knnqpv} \psi_n{\K{\knnpp,\knnqp}} \nonumber \\
  & \flb + \frac{ 1  }{\sqrt{2}} \rint{\pp} \imel{n}{p,q;\Omega_n}{O_n}{\pp,q;\Omega_n}{n} \nonumber \\
  & \flb \flb \cross \int_{-1}^{1} \dd{ \cos{\theta_{\v{\pp},\v{\qt}}} }  \cos{\theta_{\v{\pp},\v{\qt}}} \psi_c{\K{\kncp,\kncq}} \Bigg) \Bigg \rvert_{p=\knnpp,q=\knnqp}\,, \\
  d^\prime_{\np} &= 0\,.
\end{align}
As usual, the \(\kappa_{ijk}\) are functions of the momenta \(\v{\pp}\) and \(\v{\qt}\).
The vector \(\v{\qt}\) has the same absolute value as \(\v{q}\).
The only difference is now, that the whole term \(\tilde{a}^\prime_{\np}\) is evaluated at
\(p=\knnpp,q=\knnqp\) as indicated at the end of the expression.
Again, the angular integration \(\angint{\pp} \angint{\qt}\) can be simplified using the remarks
given in appendix \ref{subsec:ang_simp}.
The relation \(O_{\np} \coloneqq \K{-\pmo} O_n \K{-\pmo} \) was used in the derivation.
In this last case we deviate a bit from the definition given in Eq. (\ref{eq:def_O_i_dens}).
The index for the spectator and the one of the operator \(O\) are not equal in this case.
The correct definition for this exceptional case reads
\begin{equation}\label{eq:def_O_i_dens_np}
  O_{\np}{\K{\vpq}} \coloneqq \mel{\Psi}{\K{ \iketbra{\vpq}{n}{} \otimes \spid }O_{\np}}{\Psi}\,.
\end{equation}

The expressions for the calculation of \(O_i{\K{\vpq}}\) can be tested by considering the cases of the potentials themselves, where there is another method for
the calculation of \(V_i{\K{\vpq}}\).
From the definition of the Faddeev wave function components in Eq. (\ref{eq:def_fd_wfc}) the relation
\begin{equation}\label{eq:V_i_Psi}
  V_i \ket{\Psi} = G_0^{-1} \ket{\psi_i}
\end{equation}
follows.
Since
\begin{align}
  \imel{i}{\vpq}{G_0^{-1}}{\vpqp}{i} &= \K{E_3 - p^2/\K{2\mu_{jk}} - q^2/\K{2\mu_{i(jk)}} } \dt{\v{p}-\v{\pp}} \dt{\v{q}-\v{\qp}}
\end{align}
holds, the calculation of
\(V_i{\K{\vpq}}\) using this method requires only simple modifications of the expression for the one-body probability density
given in Eq. (\ref{eq:ob_pbd}). Note that, if  the three-body force is included as
described in subsection \ref{subsec:heft}, Eq. (\ref{eq:V_i_Psi}) is only true at vanishing three-body couplings.

We note also that, if the two-body probability density is calculated using the formulas of this section,
then for some terms transformations to a Jacobi-momentum basis corresponding to a different spectator are required.
\subsection{Results for the probability densities}\label{subsec:results}
\begin{figure}[htb!]
  \includegraphics[width=0.75\textwidth]{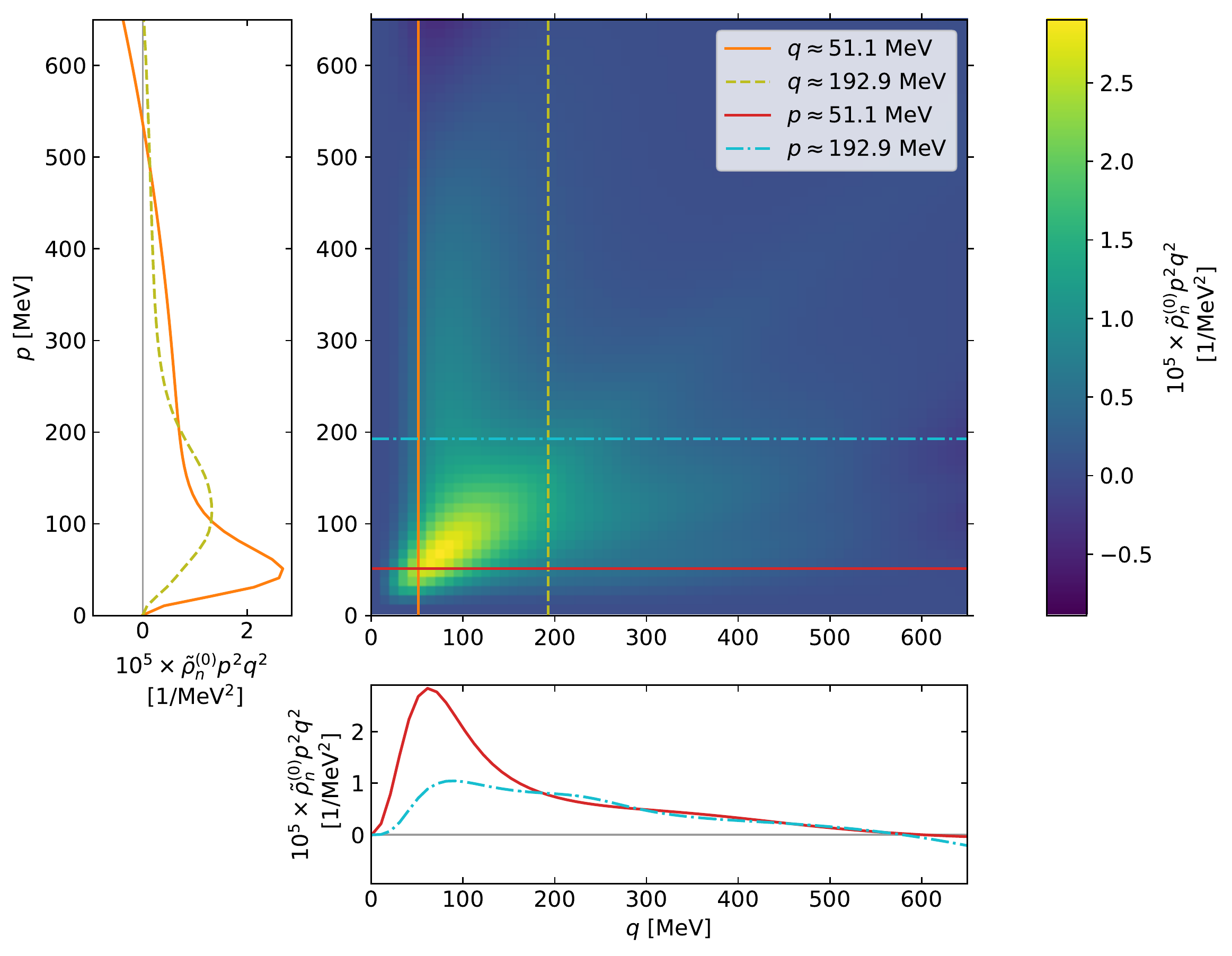}\\
  \includegraphics[width=0.75\textwidth]{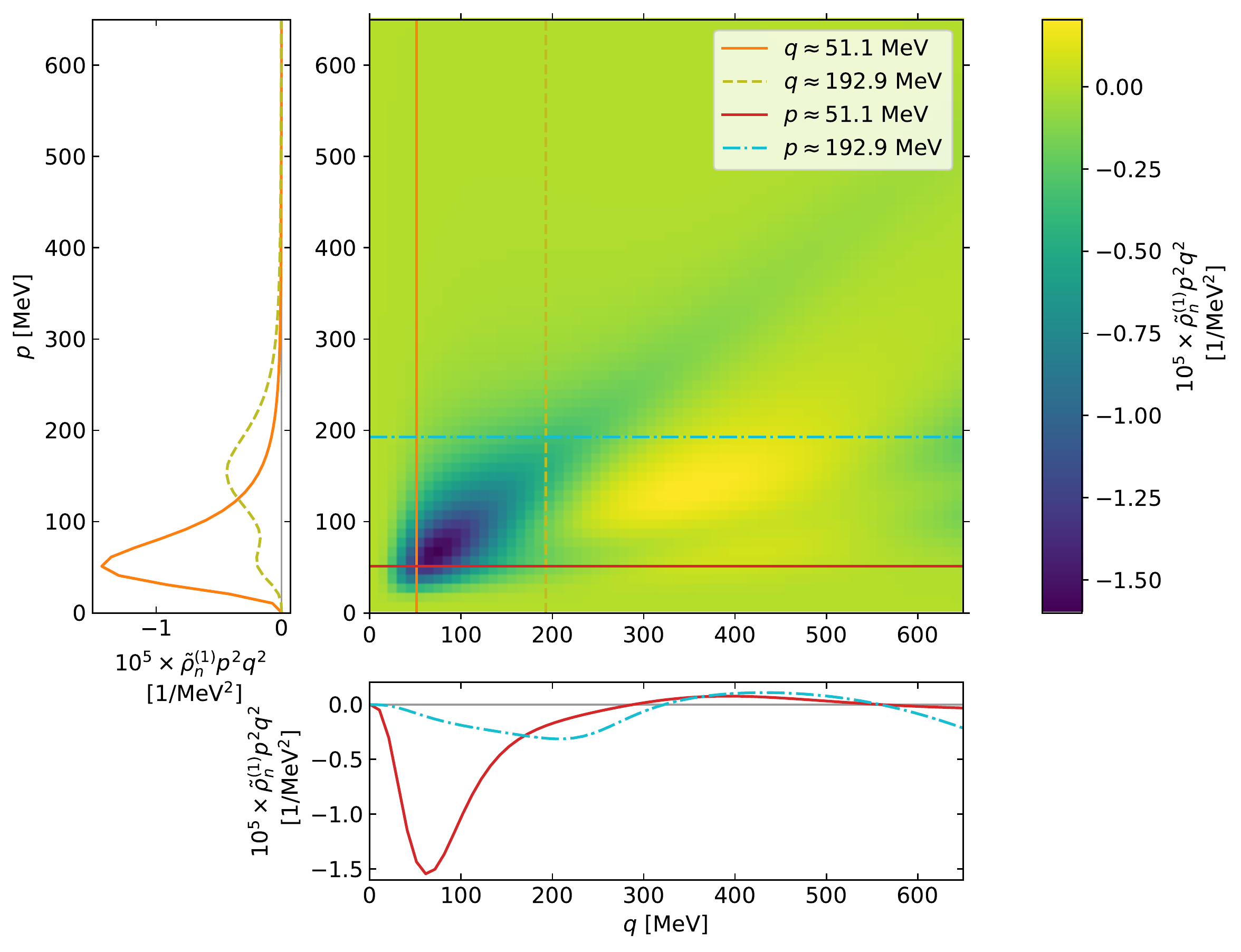}
  \caption{Projections on \(\xi=0\) (upper panel) and \(\xi=1\) (lower panel) of the approximation to the probability density \(\rhot{n}{\vpq}p^2q^2\) are shown.
  These results were obtained with \(\Lambda=750\)\,MeV. Nevertheless, we plot up to \(650\)\,MeV in order to
  show the negative regions.}\label{fig:mn_pbds}
\end{figure}

The results were obtained with the parameters \(B_3^{\K{0}} = 0.975\)\,MeV from \cite{brodeur12}, \(a_{nn} = -18.7\)\,fm from \cite{trotter06}, \(k_R = 37.4533\)\,MeV and \(r_1 = -174.0227\)\,MeV.
The latter two were calculated using the results of \cite{arndt73}.
Initially, we show results obtained with \stdff s, where we set \(\beta_0=\beta_1=\Lambda\) with the three-body cutoff \(\Lambda\).
Different angular projections done according to Eq. (\ref{eq:rhotpw}) are shown in Fig. \ref{fig:mn_pbds}.
The color plot of each of the two subfigures shows the probability density as function of the momenta \(p\)
and \(q\).
On the left and below these main plots usual 2d-plots are shown.
They depict cuts of the probability density.
The left plots show the probability density as function of \(p\) at fixed \(q\),
whereas the plots below the color plots show it as function of \(q\) at fixed \(p\).
For these fixed values a value of order (but above) the breakdown scale \(M_\mathrm{core} \approx 140\)\,MeV and
a value of order of the dineutron binding momentum \(\sqrt{2\mu_{c\K{nn}} S_{2n}} \approx \sqrt{8/3 \cross 940 \textrm{\,MeV}^2} \approx 50\)\,MeV
are chosen.
We estimate that the numerical uncertainties\footnote{
  We define the following measure for the numerical uncertainty:
  \begin{equation*}
    \frac{ \max_{i,j}{\K{\left| \rho_{ij}^{\K{2n}} - \rho_{ij}^{\K{n}} \right|}} }{ \max_{i,j}{\K{\left| \rho_{ij}^{\K{2n}} \right|}}/2 }\,,
  \end{equation*}
  where the \(\rho_{ij}\) is the probability density evaluated on a grid.
  We put a number proportional to the number of mesh points for the discretization of both the integral equations and 
 the angular integration in the brackets in the superscript of the \(\rho_{ij}\).
  These two numerical techniques---as well as others---generally use different numbers of mesh points, but the
  common proportionality factor enables us to change the overall precision.
  Accordingly this measure is the maximum absolute difference of the result with fixed numbers of mesh points and the result obtained with twice
  as many mesh points, divided by a ``typical" density value.
  For the latter we choose the half of the maximum of the absolute values.
}
of our result for the \(\xi=0\) and \(\xi=1\) projection of the (two-body) probability density are smaller than
\(1\%\).

The upper panel of Fig. \ref{fig:mn_pbds} shows that the corrections cause negative probabilities at
momenta higher than the high-momentum scale of this EFT \(M_\mathrm{core} \approx 140\)\,MeV, i.e. in regions
where we cannot trust the result anyway.
Indeed, the full probability density \eqref{eq:exactpd} is positive semi-definite, so the appearance of negative values for  \(\rhotpw{n}{0}{\vpq}p^2q^2\)
in certain regions signals that the approximation we made to evaluate it has failed there.
A contrast is provided by the lower panel of Fig. \ref{fig:mn_pbds} where \(\rhotpw{n}{1}{\vpq}p^2q^2\) is negative in the low-momentum regime.
In this case the negative values do not signal a breakdown of the approximation we made to evaluate $\tilde{\rho}$.
This projection of the probability density contains interferences between different partial waves in the wave function, and so
does not have to be non-negative.

The \(\xi=0\) projection of \(\rhot{n}{\vpq}p^2q^2\) has its peak at roughly \(p=62\)\,MeV and \(q=71\)\,MeV.
These momenta are $\sim M_{\rm halo}$ and consistent with expectations based on the $S_{2n}$ of ${}^6$He. 
The full width at half maximum (FWHM) of this peak in the \(pq\) plane is about \(150\)\,MeV.
This is consistent with $\Delta p \sim \hbar/\Delta x$ and the FWHM of  order \(2\)\,fm seen in 
the spatial probability density of \hesix~of Refs. \cite{Chulkov:1990ac,Zhukov:1993aw}.

There is an exact symmetry of \(\rhot{n}{\vpq}\) in the limit of a core-to-neutron mass ratio \(A \coloneqq m_c / m_n\) of \(\infty\):
in that limit \(\rhot{n}{\vpq}=\rhot{n}{\v{q},\v{p}}\).
The derivation of this relation is based on two ingredients.
One, true for every \(A\), is the identity \(\rhot{n}{\vpq} = \rhot{\np}{-\v{p},\v{q}}\).
This can be derived using the antisymmetry of the state \(|\Psi \rangle\) under neutron exchange and the fact that \([V_c+V_n+V_{\np} , \pmo] = 0\).
The second ingredient is the equation \(\lim_{A \to \infty} \ibraket{\np}{\vpq}{\vpqp}{n} = \dt{\v{\pp}- \v{q}} \dt{\v{\qp} + \v{p}}\).
When written in terms of magnitudes \(p\) and \(q\) the resulting relation \(\rhot{n}{\vpq}=\rhot{n}{\v{q},\v{p}}\) becomes \(\rhot{n}{p,q,\theta_{pq}}=\rhot{n}{q,p,\theta_{pq}}\).
Thus the \(\xi=0\) projection of \(\rhot{n}{\vpq}p^2q^2\)---and all higher-\(\xi\) projections too---should be approximately mirror symmetric with respect to \(p=q\).
For the case of interest to us here, \(A=4\), the symmetry seems to be quite well fulfilled by the \(\xi=0\) projection---perhaps surprisingly given that this is ``only" \(A=4\).
The symmetry is less well obeyed in the \(\xi=1\) ${}^6$He results.

Another interesting aspect of the probability density is its high-momentum tails.
There is one tail roughly parallel to the \(p\)-axis and a second roughly parallel to the \(q\)-axis.
At momenta greater than \(500\)\,MeV, where our approximation to the two-body density is definitely no longer justified, the probability density becomes negative
with calculations at higher cutoffs showing that the position in \(p\) of the zero crossing
strongly depends on the cutoff. (The parallel-to-the-\(q\)-axis tail shows  analogous behavior.)
These tails are at least outside of the region of validity of our approach, and may be pure artefacts of approximations
made in our calculation; the tails of the one-body probability density do not become negative.
The situation seems to be analogous to the example in Subsec. \ref{subsec:ve_tb_example}, where
a two-body system with an energy-dependent potential is discussed.
The one-body probability density is not normalizable, while the two-body one contains counter terms resolving this issue.
But approximations to the two-body probability density then produce cutoff-dependent zero crossings.

\begin{figure}[htb!]
  \includegraphics[width=0.49\textwidth]{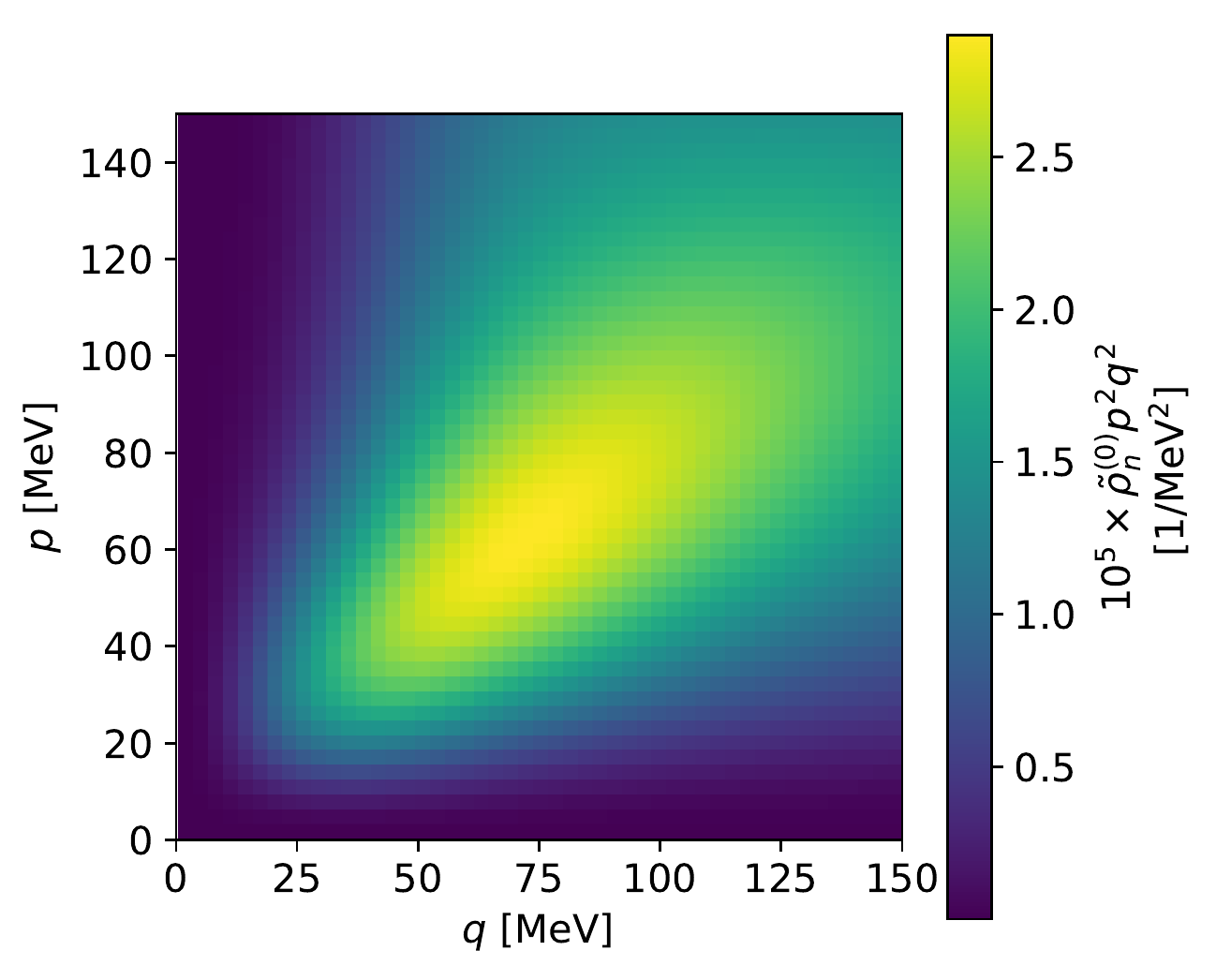}
  \includegraphics[width=0.49\textwidth]{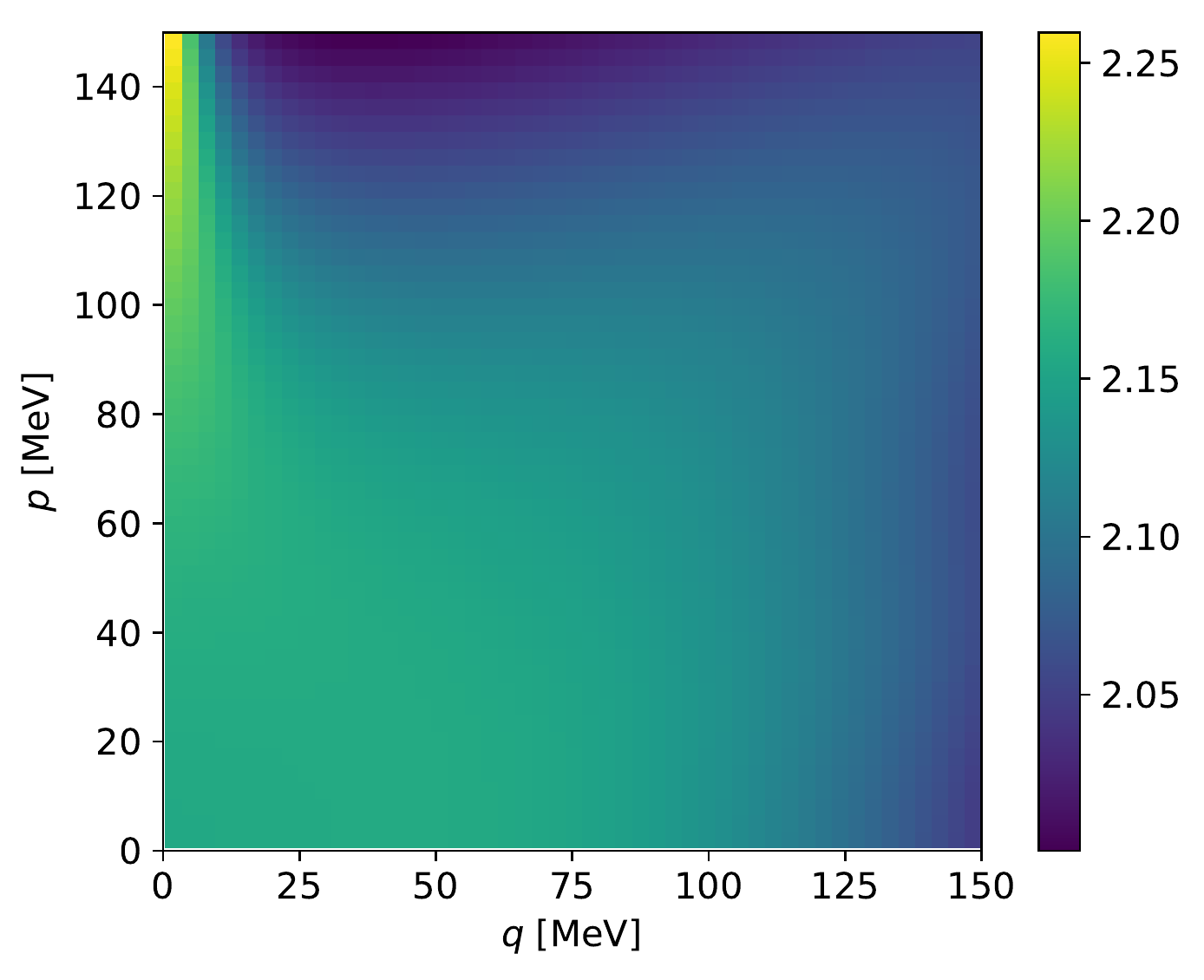}\\
  \includegraphics[width=0.49\textwidth]{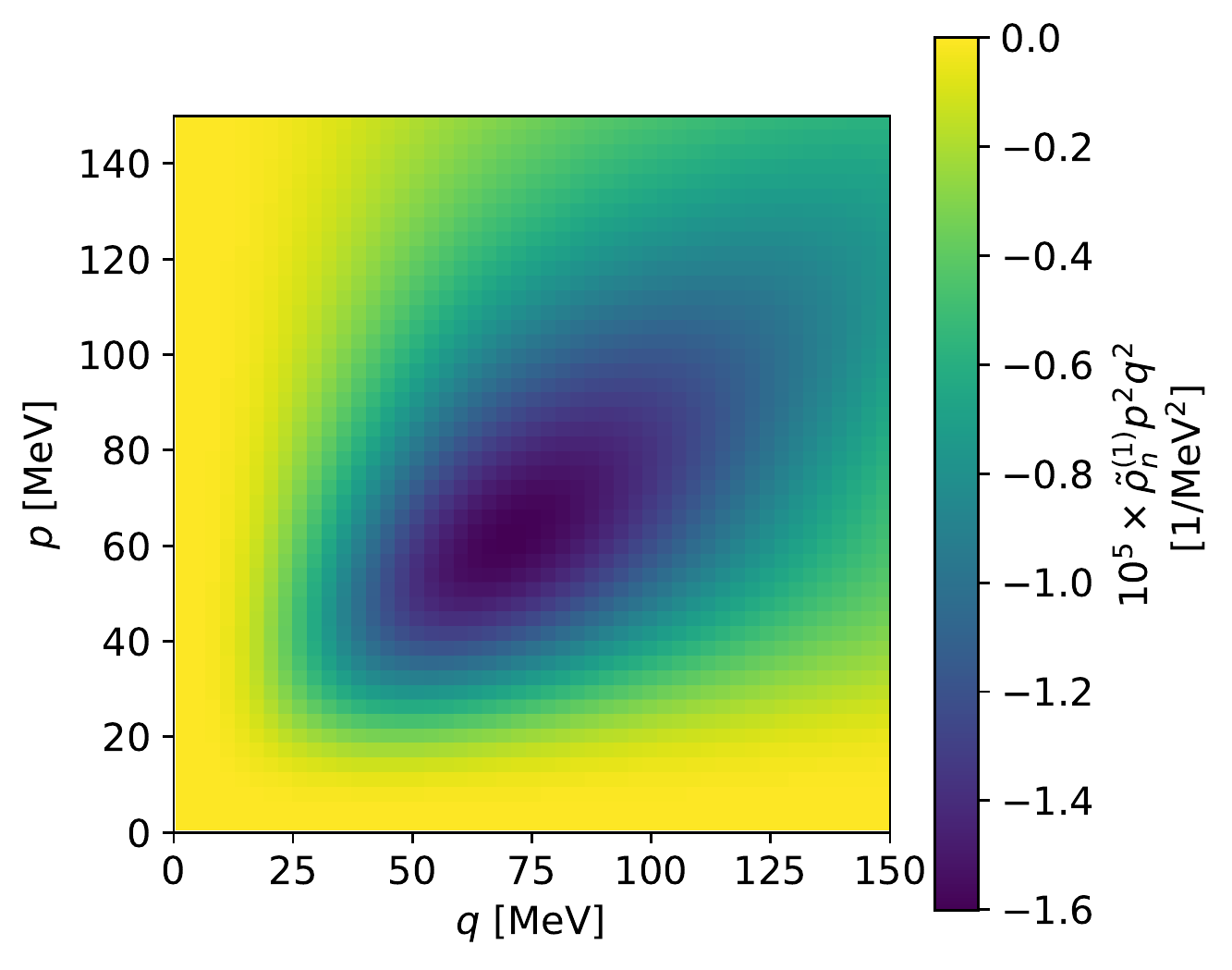}
  \includegraphics[width=0.49\textwidth]{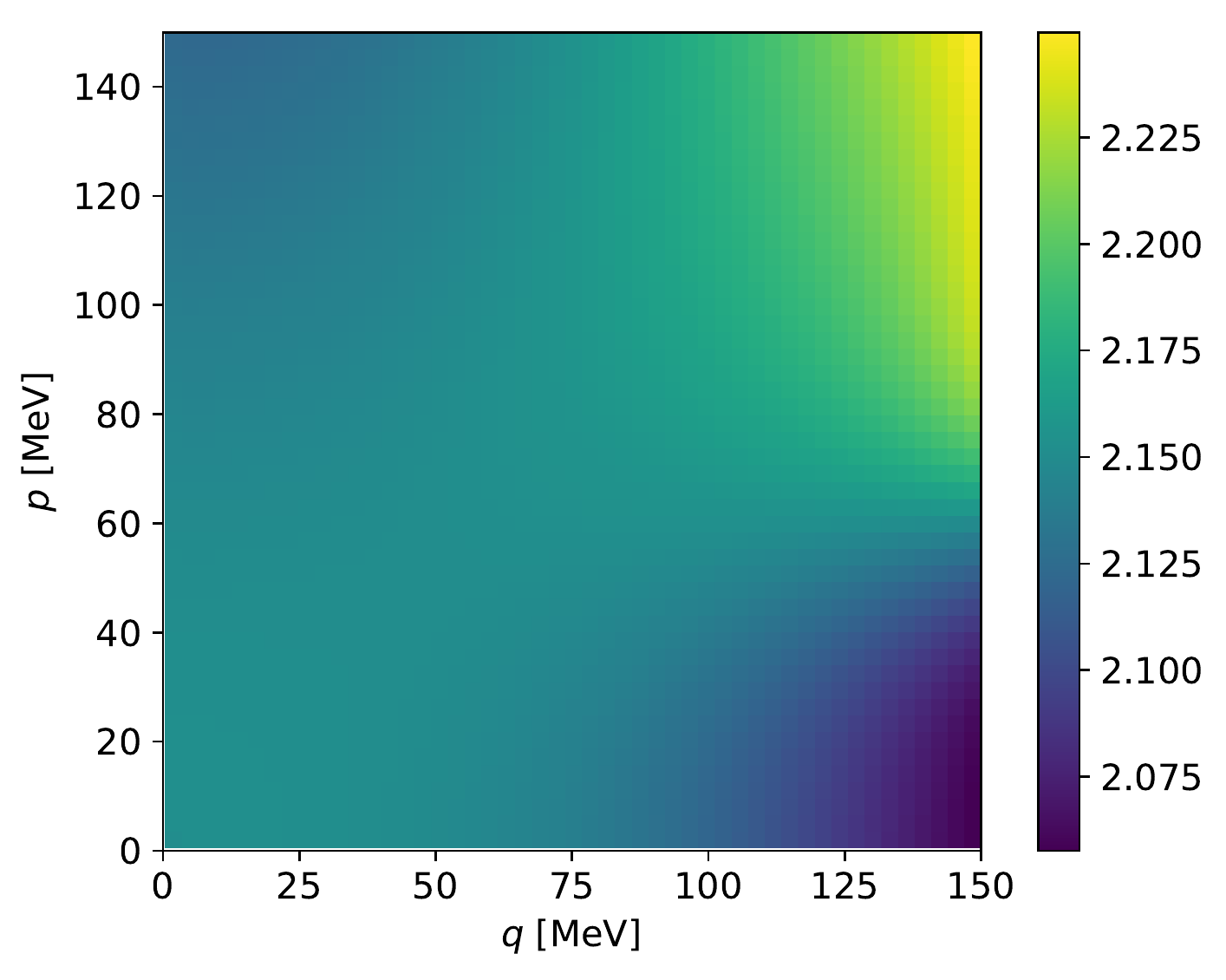}\\
  \caption{The upper left plot shows the probability density \(\rhotpw{n}{0}{p,q}p^2 q^2\), the upper right shows the quotient of probability densities \(\rhotpw{n}{0}{p,q}/\rhopw{n}{0}{p,q}\).
  The lower left plot shows \(\rhotpw{n}{1}{p,q}p^2 q^2\), the lower right shows \(\rhotpw{n}{1}{p,q}/\rhopw{n}{1}{p,q}\).
  All were obtained with \(\Lambda=750\)\,MeV.}\label{fig:pbd_div}
\end{figure}

Nevertheless, our approximation to the full density is valid at low momenta, i.e. the correction due to two-body pieces is small there, as can be seen from Fig. \ref{fig:pbd_div}, where the quotients of the
two-body and one-body probability densities are shown next to plots of the two-body probability density.
This is done for \(\xi=0\) and \(\xi=1\).
The normalization constant changes due to the modification of the probability density and contains probability values from regions where we do not trust the density anymore.
Therefore the actual values of the quotient shown in the left panel do not contain the key point.
The point is that the quotient varies very little in the domain of validity of the EFT.
Accordingly the shape of the probability density is the same with and without the two-body contributions, and so they are small in this sense.

\begin{figure}[htb!]
  \includegraphics[width=0.49\textwidth]{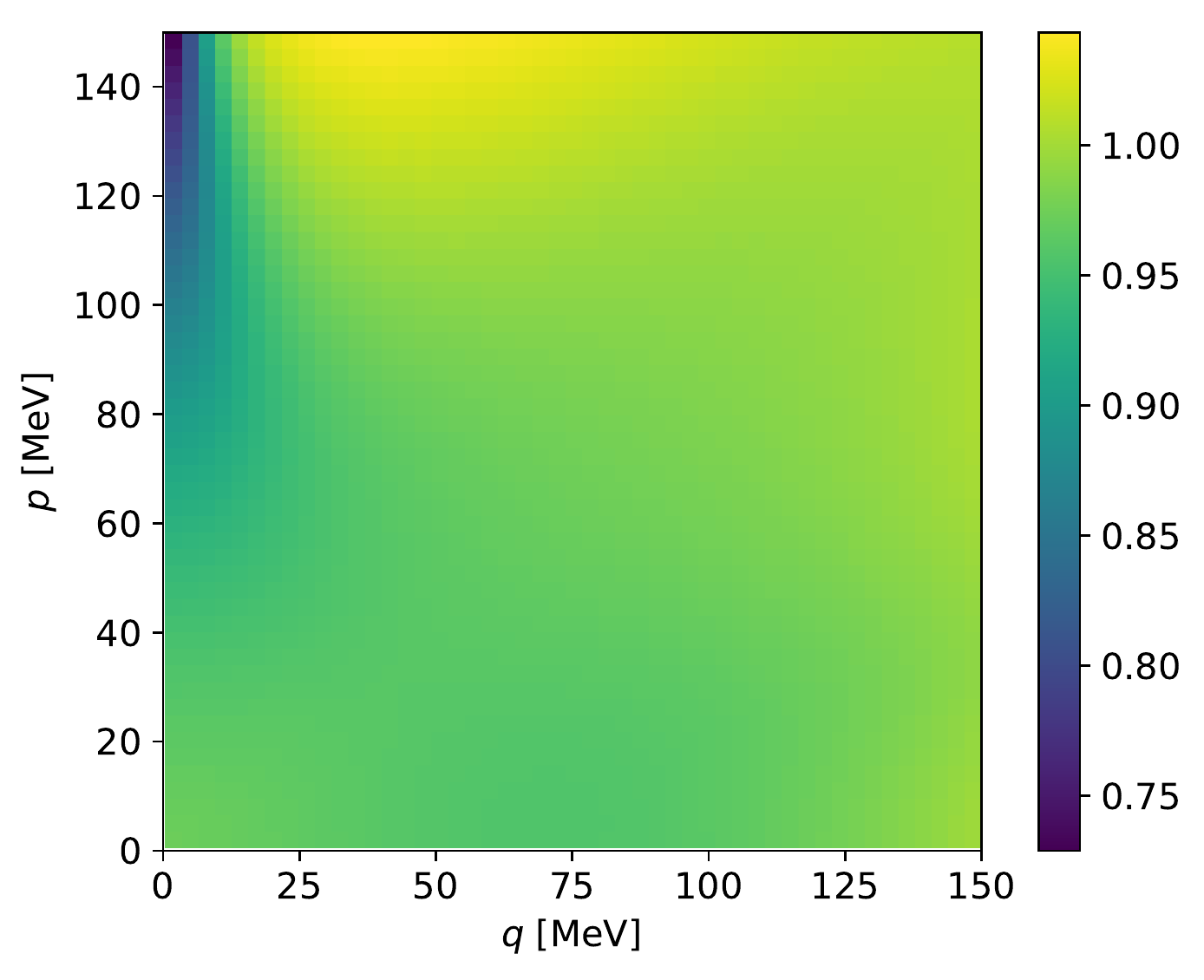}
  \includegraphics[width=0.49\textwidth]{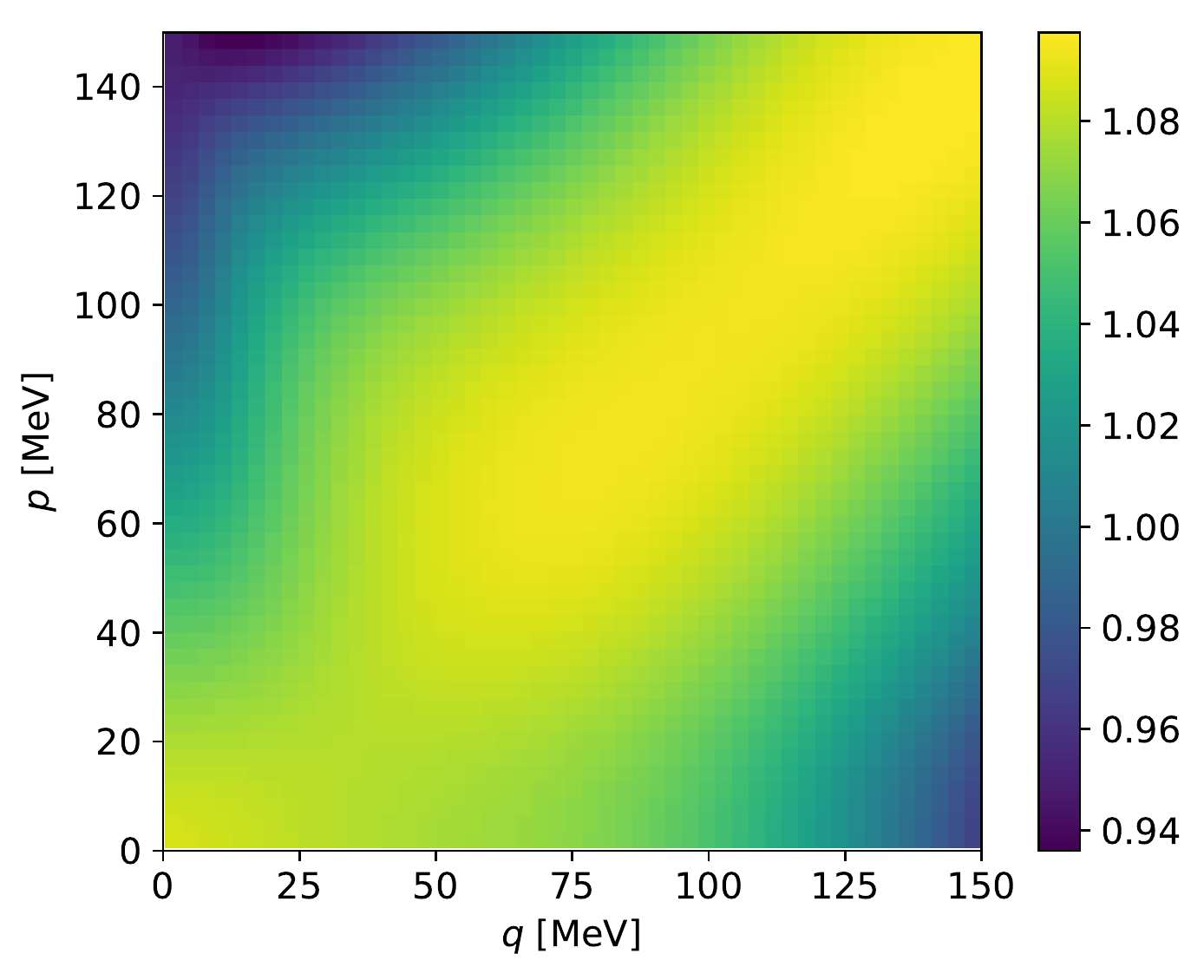}
  \caption[]{
  The left plot shows the quotient of the probability density \rhotpw{n}{0}{p,q} obtained with \(\Lambda = 1500\)\,MeV and the identical quantity obtained
  with \(\Lambda = 750\)\,MeV (both with \stdff).
  The right plot shows the quotient
  of \rhotpw{n}{0}{p,q} obtained with Yamaguchi form factors and \rhotpw{n}{0}{p,q} obtained with \stdff s.
  The right plot was obtained with \(\Lambda = 750\)\,MeV and in case of Yamaguchi form factors with \(\Lambda = 950\)\,MeV and
  \(\beta_0 = \beta_1 = 500\)\,MeV.
  It may be useful to have the plot of \(\rhotpw{n}{0}{p,q} p^2 q^2\) as given
  in the upper left of Fig. \ref{fig:pbd_div} in mind when looking at the plots of these quotients.}
  \label{fig:reg_cmp}
\end{figure}

In the left panel of Fig. \ref{fig:reg_cmp} we compare our results for the two-body probability density at low momenta for different cutoffs.
Due to the modifications resulting from the energy-dependence of the potentials
the normalization constant is in our case cutoff-dependent.
Thus again the point is that the quotient varies little in the region where the EFT is valid.
The shape of the probability density in this region is therefore, to a good approximation, cutoff independent, even though the wave function normalization is not.
This observation is also true in case of \(\rhotpw{n}{1}{p,q}p^2 q^2\).

So far, all results were obtained by using Heaviside form factors.
Finally, we compare results for the two-body probability density obtained with Yamaguchi form factors\footnote{
  Note that we make a small error by using the analytic results for the \(\lambda_i\) as given in subsection \ref{subsec:coupling_strength}, since
  in these calculations the three-body cutoff \(\Lambda\) is neglected.
} to those obtained with
Heaviside form factors in the right panel of Fig. \ref{fig:reg_cmp}.
This time, the relation between the two-body and three-body cutoffs is \(\beta_0 = \beta_1 \approx \Lambda/2\).
The shown ratio of \(\xi=0\) projections is roughly constant. This is also true for \(\xi=1\).
This shows that two different regulators in the two-body potential lead to almost the same results for the ${}^6$He wave function.

In summary, we have obtained a two-body probability density valid at low momenta.
The corrections due to the energy dependence are small in this region.
This low-momentum part is approximately regulator independent.
%
\section{Summary and Future Work}
\label{sec:future}
In this work we calculated the probability density of \hesix~in momentum space.
Within Halo Effective Field Theory, \hesix~is a three-body \(nn\ca\) bound state.
At leading order the two-body pairwise interactions are parameterized to reproduce the \(nn\) \(^1S_0\) scattering length
and the \(n\ca\) \(^2P_{3/2}\) scattering volume and effective range. A three-body force is mandatory for renormalization~\cite{ji14} and
is adjusted to reproduce the \hesix~ground-state two-neutron separation energy.

The \hesix~ground-state three-body wave function was solved in the Faddeev formalism, with the $nn$ and $n \ca$ potentials adjusted
to exactly reproduce the corresponding leading-order amplitudes of Halo EFT.
Using partial-wave decomposition, we obtained the Faddeev components, each of which contains only one spin-angular-momentum channel
in the corresponding Jacobi spectator representation. However, a single channel in one spectator representation cannot completely describe
the three-body total wave function, due to the non-trivial overlaps between spin-angular-momentum eigenstates from two different spectator representations.

The presence of energy-dependent potentials in this problem yields a modification of the
orthonormality conditions used in calculations of the momentum-space density. In addition to the
usual one-body piece of the probability density $\langle \Psi|\Psi\rangle$,
there is a new two-body term. This is linked to the energy derivative of the two-body potential,
and ensures correct normalization of the probability density when integrated over a set of momentum eigenstates.

The resulting momentum-space probability density of \hesix~is a function of the angle between the Jacobi momenta $\vec{p}$ and $\vec{q}$. We expand
this dependence in Legendre polynomials and focus mainly on the angle-averaged (zeroth moment) and $\v{p} \cdot \v{q}$ (first moment) in that expansion.
By comparing the density obtained with different regulators in the two-body potentials, we see that its low-momentum part is regulator independent.

Results for the coordinate-space probability density are desirable.
They can be obtained from the momentum-space results presented here via Fourier
transformation. But it is more numerically efficient to first transform the component functions
\(\Psi_i(p,q)\) via spherical Bessel transforms and then construct
the coordinate-space density. In either approach the transformation is numerically delicate because 
of the high-momentum behavior of the
momentum-space probability density. We postpone the presentation of results to a future publication.
 
The energy-dependent potential has been commonly used in nuclear reaction theory as well~\cite{Feshbach:1958nx}. In the scattering of nucleons from nuclei optical potentials are often introduced to account for channels of lower energy than the elastic energy, which can remove flux.
Such optical potentials are energy dependent and, strictly speaking, probabilities calculated using them should include  additional two-body terms of the type discussed here.

\begin{acknowledgements}
DRP and CJ thank Charlotte Elster for useful discussions during the early stages of this work.
DRP thanks Jerry Yang for drawing his attention to Ref.~\cite{mckellar83}.
MG thanks Wael Elkamhawy and Fabian Hildenbrand for
useful discussions during the early stages of this work.
The work of DRP was supported by the US Department of Energy under
contract DE-FG02-93ER-40756 and by the
 ExtreMe Matter Institute EMMI
 at the GSI Helmholtzzentrum f\"ur Schwerionenphysik, Darmstadt, Germany.
 HWH and MG were supported by the Deutsche Forschungsgemeinschaft (DFG, German
 Research Foundation) under project number 279384907 -- SFB 1245.
 HWH was also supported by the
 Bundesministerium f\"ur Bildung und Forschung (BMBF) through contract
 05P18RDFN1.
\end{acknowledgements}

\bibliographystyle{spphys}
\bibliography{momSpaceProbDensHe.bbl}

\appendix

\section{Alternative derivation of the normalization condition}
\label{ap:alternativederivationnorm}
In this appendix we give an alternative derivation of the normalization condition in the presence of an energy-dependent potential. This derivation focuses on the resolvent of $H(E)$. It is equivalent to the normalization conditions obtained for two-particle vertex functions in relativistic bound-state equations in Refs.~\cite{lepage77,adam97}.

The resolvent of $H(E)$ is:
\begin{equation}
G(E) \coloneqq \frac{1}{E - H(E)}\,.
\end{equation}
This contrasts with the resolvent of the energy-independent Hamiltonian obtained by evaluating $H(E)$ at $E=E_\alpha$, $\bar{H} \coloneqq H(E_\alpha)$. Denoting that resolvent by $\bar{G}(E)$ we have:
\begin{equation}
\bar{G}(E) \coloneqq \frac{1}{E-\bar{H}}\,.
\end{equation}
By construction, $\bar{H}$ and $H(E)$ both have $|\psi_\alpha \rangle$ as a right eigenstate corresponding to energy $E_\alpha$. The rest of $\bar{H}$'s spectrum will, however, be different. And, while the eigenstate $|\psi_\alpha \rangle$ is common to the spectrum of both operators, it is not guaranteed that it should be normalized in the same way. Indeed, we will show here that it should not be.

Expanding $H(E)$ around $E=E_\alpha$ yields:
\begin{equation}
G(E)=\frac{1}{E-\bar{H} - (E-E_\alpha) V'(E_\alpha) - \ldots}\,,
\end{equation}
where $'$ denotes differentiation with respect to $E$.
This, in turn, allows us to write:
\begin{equation}
G(E)=\bar{G}(E)\left\{1 + \sum_{n=1}^\infty [V'(E_\alpha) (E-E_\alpha) \bar{G}(E)]^n\right\} + \mbox{regular as $E \rightarrow E_\alpha$}\,.
\label{eq:resumming}
\end{equation}
Since $\bar{G}(E)$ is the resolvent of  an energy-independent Hermitian operator it has a standard spectral representation. If $E_\alpha$ is an isolated bound state then that is:
\begin{equation}
\bar{G}(E)=\frac{|\bar{\psi}_\alpha \rangle \langle \bar{\psi}_\alpha|}{E-E_\alpha} + \mbox{pieces in the  space orthogonal to $| \bar{\psi}_\alpha \rangle$}\,.
\label{eq:spectral}
\end{equation}
where we have used $|\bar{\psi}_\alpha \rangle$ to emphasize that this eigenstates of $\bar{H}$ obey the standard normalization condition $\langle \bar{\psi}_\beta|\bar{\psi}_\alpha \rangle=\delta_{\beta \alpha}$, since $\bar{H}$ is an energy-independent, Hermitian, operator. As discussed at length in Sec.~\ref{sec:energydeppotentials} the states
$|\psi_\alpha \rangle$ do not obey this condition, although for $\beta \neq \alpha$ this is not a surprise, since the spectrum of $\bar{H}$ differs from that of $H$ apart from the one state at $E=E_\alpha$. But for that particular state we must write:
\begin{equation}
{\cal Z}^{1/2} |\psi_\alpha \rangle=|\bar{\psi}_\alpha \rangle\,,
\end{equation}
with ${\cal Z}$ an additional wave function renormalization associated with the energy dependence of $V(E)$.

To fix ${\cal Z}$ we
insert Eq.~(\ref{eq:spectral}) in Eq.~(\ref{eq:resumming}) and sum the geometric series. This yields:
\begin{equation}
G(E)=\frac{|\bar{\psi}_\alpha \rangle \langle \bar{\psi}_\alpha|}{E-E_\alpha} \frac{1}{1 - \langle \bar{\psi}_\alpha| V'(E_\alpha))|\bar{\psi}_\alpha \rangle} +  \mbox{regular as $E \rightarrow E_\alpha$}\,.
\label{eq:GofEspectral}
\end{equation}
The factor in the denominator here ensures that the consistency condition:
\begin{equation}
G'(E)=-G(E) \left[\frac{\partial}{\partial E} G^{-1}(E)\right] G(E)\,.
\end{equation}
is satisfied.
The spectral decomposition (\ref{eq:GofEspectral}) will then take the standard form:
\begin{equation}
G(E)=\frac{|\psi_\alpha \rangle \langle \psi_\alpha|}{E-E_\alpha} +  \mbox{regular as $E \rightarrow E_\alpha$}\,,
\end{equation}
provided we identify:
\begin{equation}
{\cal Z}=1 - \langle \bar{\psi}_\alpha| V'(E_\alpha)|\bar{\psi}_\alpha\rangle=\langle \bar{\psi}_\alpha|\id - V'(E_\alpha)|\bar{\psi}_\alpha \rangle\,.
\label{eq:Z}
\end{equation}

In practice this means that we solve the Hamiltonian eigenvalue problem at $E=E_\alpha$, thereby obtaining an eigenvector that is, prior to normalization, proportional to both $|\bar{\psi}_\alpha \rangle$ and $|\psi_\alpha \rangle$. Then, if we wish to compute the latter, we must normalize such that:
\begin{equation}
\langle \psi|\id - V'(E_\alpha)|\psi \rangle=1\,,
\end{equation}
instead of using the standard
\begin{equation}
\langle \psi|\psi \rangle=1\,.
\end{equation}
%
\section{Additional formulas for the probability density}
\subsection{Transformation of Jacobi coordinates}\label{subsec:jacobi}
For the evaluation of overlaps of the type \(\ibraket{i}{\vpq}{\vpqp}{j}\) the following functions are useful, since
they abbreviate many expressions:
\begin{align}
  \pif{1}{\v{p},\v{q}} &\coloneqq \v{p} + \frac{A}{A+1} \v{q}\,,\\
  \pif{2}{\v{p},\v{q}} &\coloneqq \v{p} + \frac{1}{2} \v{q}\,,\\
  \pif{3}{\v{p},\v{q}} &\coloneqq \v{p} + \frac{1}{A+1} \v{q}\,.
\end{align}
The transformations can be summarized by the expression
\begin{equation}
  \ibraket{i}{\vpq}{\vpqp}{j} \eqqcolon \dt{\v{\pp}-\kappa_{ijp}{\K{\vpq}}} \, \dt{\v{\qp}-\kappa_{ijq}{\K{\vpq}}} \,,
\end{equation}
which defines \(\v{\kappa}_{ijk}{\K{\vpq}}\) with \(i,j \in \{n,c\}\), \(i \neq j\) and \(k \in \{p,q\}\).
The functions \(\v{\kappa}_{ijk}\) are given by
\begin{align}
  \kncpv &\coloneqq \pfv{2}{\v{q}, -\pfv{1}{\v{p},\v{q}}}\,, \\
  \kncqv &\coloneqq -\pfv{1}{\v{p},\v{q}}\,, \\
  \kcnpv &\coloneqq -\pfv{1}{\v{q},\pfv{2}{\v{p},-\v{q}}}\,, \\
  \kcnqv &\coloneqq \pfv{2}{\v{p},-\v{q}}\,.
\end{align}
Furthermore, we define the functions \(\v{\kappa}^\prime_{ijk}{\K{\vpq}}\) with \(i,j \in \{n,c\}\) and \(k \in \{p,q\}\) by
\begin{equation}
  \imel{i}{\vpq}{\pmospatial}{\vpqp}{j} \eqqcolon \dt{\v{\pp}-\v{\kappa}^\prime_{ijp}{\K{\vpq}}} \, \dt{\v{\qp}-\v{\kappa}^\prime_{ijq}{\K{\vpq}}} \,,
\end{equation}
where the spatial part of the \(nn\) permutation operator \(\pmo\) is given by \(\pmospatial\).
The used functions of the type \(\v{\kappa}^\prime_{ijk}\) read
\begin{align}
  \knnpv^\prime &\coloneqq \pfv{3}{\v{q},\pfv{3}{\v{p},-\v{q}}}\,, \\
  \knnqv^\prime &\coloneqq \pfv{3}{\v{p},-\v{q}}\,, \\
  \kcnpv^\prime &\coloneqq -\pfv{1}{\v{q},-\pfv{2}{\v{p},\v{q}}}\,, \\
  \kcnqv^\prime &\coloneqq -\pfv{2}{\v{p},\v{q}}\,.
\end{align}
\subsection{Simplification of certain combinations of coupled spherical harmonics}\label{subsec:ang_simp}
In this appendix identities simplifying the angular dependence of the probability density are given.
The angle \(\gamma\) between two vectors \(\v{v}_1\) and \(\v{v}_2\) is defined by
\begin{align}
  \cos{\gamma\K{\theta_1, \varphi_1, \theta_2, \varphi_2 }} \coloneqq \v{\myhat{v}_1} \cdot \v{\myhat{v}_2}\, \\
  \v{\myhat{v}_i} = \begin{pmatrix} \sin{\theta_i} \cos{\varphi_i} \\ \sin{\theta_i} \sin{\varphi_i} \\ \cos{\theta_i} \end{pmatrix}\,.
\end{align}
Expressed in terms of the angles of the two vectors it reads
\begin{equation}\label{eq:cos_gamma}
  \cos{\gamma\K{\theta_1,\varphi_1,\theta_2,\varphi_2}} \coloneqq
    \cos{\theta_1} \cos{\theta_2} + \sin{\theta_1} \sin{\theta_2} \cos{\K{\varphi_1 - \varphi_2}}\,.
\end{equation}
In fact, the angular dependence of the coupled spherical harmonic \(\cywa{11}{00}\) is captured entirely by this angle \(\gamma\).
\begin{equation}\label{eq:csh_11_00_s}
  \cy{11}{00}{\theta_1, \varphi_1, \theta_2, \varphi_2} = \frac{-\sqrt{3}}{4\pi} \cos{\gamma\K{\theta_1, \varphi_1, \theta_2, \varphi_2}}\,.
\end{equation}
Also the following combination of coupled spherical harmonics can be purely expressed in terms of relative angles:
\begin{align}\label{eq:csh_11_sum_s}
  & \sum_{M=-1}^{1} \cy{11}{1M}{\theta_1, \varphi_1, \theta_2, \varphi_2} \K{\cy{11}{1M}{\theta_1^\prime, \varphi_1^\prime, \theta_2^\prime, \varphi_2^\prime}}^* \nonumber \\
  & \flb = \frac{1}{2} \K{ \frac{3}{4\pi}}^2 \bigg( \cos{\gamma\K{\theta_1, \varphi_1, \theta_1^\prime, \varphi_1^\prime }} \cos{\gamma\K{\theta_2, \varphi_2, \theta_2^\prime, \varphi_2^\prime}} \nonumber \\
  & \flb - \cos{\gamma\K{\theta_1, \varphi_1, \theta_2^\prime, \varphi_2^\prime }} \cos{\gamma\K{\theta_1^\prime, \varphi_1^\prime, \theta_2, \varphi_2}} \bigg) \,.
\end{align}
Although these relative angles are angles between different vectors, their use simplifies the probability density a lot.
If the identity is applied to our probability density all these relative angles will be only functions of \(p\), \(q\) and \(\theta_{pq}\).
Therefore angular integrations over the probability density simplify significantly.
Instead of four angular integrals only the one over \(\theta_{pq}\) has to be carried out numerically.
\subsection{Numerical calculation of \(X_{ij}\)}\label{subsec:num_Xij}
In the following expressions for the numerical calculation of the \(X_{ij}\) are given for any type of form factor.
In a first step the expression for \(X_{nn}\) is derived using the identity
\begin{align}\label{eq:Xnn_id}
  &\imel{n}{p,q;\Omega_n}{-\pmo}{\pp,\qp;\Omega_n}{n} = -\angint{p} \angint{q} \angint{\pp} \angint{\qp} \nonumber \\
  &\flbc \imel{n}{p,q;\Omega_n}{ \K{ \iket{\vpq \vphantom{\vpqp} }{n} \imel{n}{\vpq}{\pmospatial}{\vpqp}{n} \ibra{n}{\vpqp} \otimes \pmospin } }{\pp,\qp;\Omega_n}{n}\,.
\end{align}
The calculation yields
\begin{align}
  &X_{nn}{\K{q,\qp;E}} \nonumber \\
  & \flb = \rint{\p} \rint{\pp} \reg{n}{p} \gz{n}{p,q;E} \reg{n}{\pp} \imel{n}{p,q;\Omega_n}{-\pmo }{\pp,\qp;\Omega_n}{n} \nonumber \\
  & \flb = - \angint{p} \angint{q} \angint{\pp} \angint{\qp} \reg{n}{\pifnv{3}{\v{\qp},\v{\q}}} \gz{n}{\pifnv{3}{\v{\qp},\v{\q}}, q; E} \reg{n}{\pifnv{3}{\v{\q},\v{\qp}}} \nonumber \\
  & \flb \flbc \sum_{L=0}^1 \sum_{M=-L}^L \frac{ \K{-2}^{1-L} }{ 6L+3 } \K{ \cy{11}{L M}{\vpqh} }^* \cy{11}{L M}{\vpqph} \nonumber \\
  & \flb \flbc \da{\vh{p} - \pifvh{3}{\v{\qp}, \v{q}}} \da{\vh{\pp} - \pifvh{3}{\v{q}, \v{\qp}}} \nonumber \\
  & \flb = - \angint{q} \angint{\qp} \reg{n}{\pifnv{3}{\v{\qp},\v{\q}}} \gz{n}{\pifnv{3}{\v{\qp},\v{\q}}, q; E} \reg{n}{\pifnv{3}{\v{\q},\v{\qp}}} \nonumber \\
  & \flb \flbc \sum_{L=0}^1 \sum_{M_L=-L}^L \frac{ \K{-2}^{1-L} }{ 6L+3 } \K{ \cy{11}{L M_L}{\pif{3}{\v{\qp}, \v{q}}, \v{q}} }^* \cy{11}{L M_L}{\pif{3}{\v{q}, \v{\qp}}, \v{\qp}} \,,
\end{align}
where \(\da{\hat{\v{p}}- \hat{\v{\pp}}} \coloneqq \de{\varphi-\varphi^\prime} \frac{\de{\theta-\theta^\prime}}{\sin{\theta}}\) holds.
In order to evaluate this expression only one angular integral has to be computed numerically by using the identities given in appendix \ref{subsec:ang_simp}

The calculation of \(X_{nc}\) is based on an identity similar to Eq. (\ref{eq:Xnn_id}):
\begin{align}\label{eq:Xnc_id}
  &\ibraket{n}{p,q;\Omega_n}{\pp,\qp;\Omega_c}{c} = \angint{p} \angint{q} \angint{\pp} \angint{\qp} \nonumber \\
  &\flbc \imel{n}{p,q;\Omega_n}{ \K{ \iket{\vpq \vphantom{\vpqp} }{n} \ibraket{n}{\vpq}{\vpqp}{c} \ibra{c}{\vpqp} \otimes \spid } }{\pp,\qp;\Omega_c}{c}\,.
\end{align}
With it one obtains
\begin{align}
  &X_{nc}{\K{q,\qp;E}} \nonumber \\
  & \flb = \rint{\p} \rint{\pp} \reg{n}{p} \gz{n}{p,q;E} \reg{c}{\pp} \ibraket{n}{p,q;\Omega_n}{\pp,\qp;\Omega_c}{c} \nonumber \\
  & \flb = - \angint{p} \angint{q} \angint{\pp} \angint{\qp} \reg{n}{\pifnv{1}{\v{\qp},\v{\q}}} \gz{n}{\pifnv{1}{\v{\qp},\v{\q}}, q; E} \reg{c}{\pifnv{2}{\v{\q},\v{\qp}}} \nonumber \\
  & \flb  \flbc \sqrt{\frac{2}{3}} \K{ \cy{11}{00}{\vpq} }^* \cy{00}{00}{\v{\pp}, \v{\qp}} \da{\vh{p} + \pifvh{1}{\v{\qp}, \v{q}}}
  \da{\vh{\pp} - \pifvh{2}{\v{q}, \v{\qp}}} \nonumber \\
  & \flb = - \angint{q} \angint{\qp} \reg{n}{\pifnv{1}{\v{\qp},\v{\q}}} \gz{n}{\pifnv{1}{\v{\qp},\v{\q}}, q; E} \reg{c}{\pifnv{2}{\v{\q},\v{\qp}}} \nonumber \\
  & \flb  \flbc \sqrt{\frac{2}{3}} \K{ \cy{11}{00}{-\pif{1}{\v{\qp}, \v{q}}, \v{q}} }^* \cy{00}{00}{\pif{2}{\v{q}, \v{\qp}}, \v{\qp}} \nonumber \\
  & \flb = \frac{1}{\sqrt{2}} \int_{-1}^{1} \dd{\cos{\theta_{q,\qp}}} \reg{n}{\pifnv{1}{\v{\qp},\v{\q}}} \gz{n}{\pifnv{1}{\v{\qp},\v{\q}}, q; E} \reg{c}{\pifnv{2}{\v{\q},\v{\qp}}}
  \cos{\theta_{-\pif{1}{\v{\qp}, \v{q}},\v{q}}} \,,
\end{align}
where the definition \(\theta_{\v{a},\v{b}} \coloneqq \arccos{ \K{\v{a} \cdot \v{b}/\K{ab}}}\) holds.
Note that
\begin{equation}
  X_{nc}{\K{q,\qp;E}} = X_{cn}{\K{\qp,q;E}}
\end{equation}
holds. This can be shown using that \(\ibraket{n}{p,q;\Omega_n}{\pp,\qp;\Omega_c}{c}\) is real.

The implementation of these results was checked by evaluating the expressions for \stdff s, for which the analytic expressions have been calculated in \cite{ji14}.
As in the derivation of the analytic expressions the \stdff s~have been neglected, discrepancies from the numerical results can occur at momenta of the order of the regularization scale.
\subsection{Overlaps in partial wave basis}\label{subsec:ov_pwb}
In the following expressions for the overlaps of partial wave states with different spectators are derived.
Also matrix elements of the \(nn\) permutation operator \(\pmo = \pmospatial \otimes \pmospin\) are given in this basis.
We obtain
\begin{align}
  & \ibraket{c}{p,q;\Omega_c}{\pp,\qp;\Omega_n}{n} \nonumber \\
  & \flb = \angint{p} \angint{q} \angint{\pp} \angint{\qp} \nonumber \\
  & \flb \flb \cross \imel{c}{p,q;\Omega_c}{ \K{ \iket{\vpq \vphantom{\vpqp} }{c} \ibraket{c}{\vpq}{\vpqp}{n} \ibra{n}{\vpqp} \otimes \spid } }{\pp,\qp;\Omega_n }{n} \nonumber \\
  & \flb = \frac{ \sqrt{2} }{\K{4\pi}^2} \angint{p} \angint{q} \angint{\pp} \angint{\qp} \nonumber \\
  & \flb \flb \cross \cos{\theta_{\vpqp}} \dt{ \v{\pp} - \kcnpv{\K{\vpq}} } \dt{ \v{\qp} - \kcnqv{\K{\vpq}} }\,, \label{eq:cOn_aux} \\
%
%
  & \imel{c}{p,q;\Omega_c}{\pmo}{\pp,\qp;\Omega_n}{n} \nonumber \\
  & \flb = \angint{p} \angint{q} \angint{\pp} \angint{\qp} \nonumber \\
  & \flb \flb \cross \imel{c}{p,q;\Omega_c}{ \K{ \iket{\vpq \vphantom{\vpqp} }{c} \imel{c}{\vpq}{\pmospatial}{\vpqp}{n} \ibra{n}{\vpqp} \otimes \pmospin } }{\pp,\qp;\Omega_n }{n} \nonumber \\
  & \flb = -\frac{ \sqrt{2} }{\K{4\pi}^2} \angint{p} \angint{q} \angint{\pp} \angint{\qp} \nonumber \\
  & \flb \flb \cross \cos{\theta_{\vpqp}} \dt{ \v{\pp} - \kcnppv{\K{\vpq}} } \dt{ \v{\qp} - \kcnqpv{\K{\vpq}} }\,, \label{eq:cOn_aux_wp} \\
  & \ibraket{n}{p,q;\Omega_n}{\pp,\qp;\Omega_c}{c} \nonumber \\
  & \flb = \angint{p} \angint{q} \angint{\pp} \angint{\qp} \nonumber \\
  & \flb \flb \cross \imel{n}{p,q;\Omega_n}{ \K{ \iket{\vpq \vphantom{\vpqp} }{n} \ibraket{n}{\vpq}{\vpqp}{c} \ibra{c}{\vpqp} \otimes \spid } }{\pp,\qp;\Omega_c }{c} \nonumber \\
  & \flb = \frac{\sqrt{2}}{\K{4\pi}^2} \angint{p} \angint{q} \angint{\pp} \angint{\qp} \nonumber \\
  & \flb \flb \cross \cos{\theta_{\vpq}} \dt{ \v{\pp} - \kncpv{\K{\vpq}} } \dt{ \v{\qp} - \kncqv{\K{\vpq}} }\,, \label{eq:nOc_aux} \\
  & \imel{n}{p,q;\Omega_n}{\pmo}{\pp,\qp;\Omega_n}{n} \nonumber \\
  & \flb = \angint{p} \angint{q} \angint{\pp} \angint{\qp} \nonumber \\
  & \flb \flb \cross \imel{n}{p,q;\Omega_n}{ \K{ \iket{\vpq \vphantom{\vpqp} }{n} \imel{n}{\vpq}{\pmospatial}{\vpqp}{n} \ibra{n}{\vpqp} \otimes \pmospin } }{\pp,\qp;\Omega_n }{n} \nonumber \\
  & \flb = \angint{p} \angint{q} \angint{\pp} \angint{\qp}
  \sum_{L=0}^1 \sum_{M=-L}^{L} \K{-1}^{1-L} \frac{2^{1-L}}{6L+3} \nonumber \\
  & \flb \flb \cross \K{\cy{11}{LM}{\v{p},\v{q}}}^* \cy{11}{LM}{\vpqp} \dt{ \v{\pp} - \knnppv{\K{\vpq}} } \dt{ \v{\qp} - \knnqpv{\K{\vpq}}}\,. \label{eq:nOn_aux_wp}
\end{align}

\end{document}